\documentstyle[12pt,epsfig,doublespace]{article}
\begin{document}
\begin{titlepage}
\begin{flushright}
ADP-00-44$/$T427 \\
PCCF-RI-0018
\end{flushright}
%%%%%%%%%%%%%%%%%%%%%%%%%%%%%%%%%%%%%%%%%%%%%%%%%%%%%%%%%%%%%%%%%%%%%%%%%%%%%
\renewcommand{\thefootnote}{\fnsymbol{footnote}}
\vspace{-0.5em}
\begin{center}
{\LARGE Enhanced Direct CP Violation in $B^{\pm} \rightarrow \rho^{0} \pi^{\pm}$ }
% 
%{\LARGE Direct CP Violation  of B Meson via $\rho-\omega$ Mixing and Range $N_{c}$}
%
\end{center}
\begin{center}
\begin{large}
X.-H. Guo$^{1}$\footnote{xhguo@physics.adelaide.edu.au}, 
O. Leitner$^{1,2}$\footnote{oleitner@physics.adelaide.edu.au}, 
A.W. Thomas$^1$\footnote{athomas@physics.adelaide.edu.au} \\
\end{large}
$^1$ Department of Physics and Mathematical Physics, and \\
Special Research Center for the Subatomic Structure of Matter, \\
Adelaide  University , Adelaide 5005, Australia \\
$^2$ Laboratoire de Physique Corpusculaire, Universit\'e Blaise Pascal, \\
CNRS/IN2P3, 24 avenue des Landais, 63177 Aubi\`ere Cedex, France 
\end{center}
\vspace{-1.5em}
\begin{abstract}
We study direct CP violation in the hadronic decay $B^{\pm} \rightarrow \rho^{0}\pi^{\pm}$, including the effect of   $\rho - \omega$ mixing. We find that the CP violating asymmetry is strongly dependent on the  CKM matrix elements, especially the Wolfenstein parameter $\eta$. For  fixed $N_{c}$ (the effective parameter associated with factorization), the CP violating asymmetry, $a$, has a maximum  of order $30\%-50\%$  when the invariant mass of the $\pi^{+}\pi^{-}$ pair is in the vicinity of the $\omega$ resonance. The sensitivity of the asymmetry, $a$, to $N_{c}$ is small. Moreover,  if $N_{c}$ is constrained using  the latest  experimental branching ratios  from the CLEO collaboration, we find that the sign of $\sin \delta$  is always positive. Thus, a measurement of direct CP violation in $B^{\pm}  \rightarrow \rho^{0}\pi^{\pm}$ would  remove the mod$(\pi)$ ambiguity in ${\rm arg}\left[ - \frac{V_{td}V_{tb}^{\star}}{V_{ud}V_{ub}^{\star}}\right]$. 
%pu
%
\end{abstract}
\vspace{-1em}
\newline
%
%
%
%Keywords: CP violation, B-decay, form factor, CKM matrix element, branching ratio
%
%%%%%%%%%%%%%%%%%%%%%%%%%%%%%%%%%%%%%%%%%%%%%%%%%%%%%%%%%%%%%%%%%%%%%%%%%%%%%
\end{titlepage}
\newpage

\section{Introduction}
Even though CP violation has been known since  $1964$, we still do not know  the source of CP violation clearly. In the Standard Model, a non-zero phase angle in the  Cabbibo-Kobayashi-Maskawa (CKM) matrix  is responsible for  CP violating  phenomena.  In the past few years, numerous theoretical studies have been conducted on  CP violation in the B meson system~\cite{ref1,ref2}. However, we need   a lot of data to check these  approaches because  there are many theoretical uncertainties  -- e.g.  CKM matrix elements, hadronic matrix elements and nonfactorizable effects.  The future aim would be to reduce  all  these uncertainties.
\newline
Direct  CP violating asymmetries in B decays  occur  through the interference of at least two amplitudes with different weak  phase $\phi$ {\em and} strong phase $\delta$. In order to extract  the weak phase   (which is determined by the CKM matrix elements),  one must know  the strong phase   $\delta$ and this is usually not well determined. In addition, in order to have a large signal, we have to appeal to  some phenomenological mechanism to obtain  a large $\delta$. The charge symmetry violating mixing between $\rho^{0}$ and $\omega$, can be extremely important in this regard. In particular, it can lead to   a  large CP violation in B decays such as $B^{\pm} \rightarrow \rho^{0}(\omega) \pi^{\pm} \rightarrow \pi^{+} \pi^{-}  \pi^{\pm} $, because  the strong phase  passes  through $90^{o}$ at the $\omega$ resonance~\cite{ref3,ref4,ref5}. Recently, CLEO reported new data~\cite{ref6} on $B \rightarrow \rho \pi$. It is the aim  of the present work to analyse direct CP violation in $B^{\pm} \rightarrow \rho^{0}(\omega) \pi^{\pm}\rightarrow \pi^{+} \pi^{-} \pi^{\pm}$, including   $\rho-\omega$ mixing, using   the latest data from the CLEO collaboration to constrain the calculation. In order to extract the strong phase $\delta$, we  use the factorization approach, in which the hadronic matrix elements of operators are saturated by vacuum intermediate states. 
\newline
In this paper, we  investigate five  phenomenological models with different weak form factors and  determine the CP violating asymmetry for $B^{\pm} \rightarrow \rho^{0}(\omega) \pi^{\pm} \rightarrow \pi^{+} \pi^{-}\pi^{\pm} $  in these models. We select models which are consistent with the CLEO data and  determine the  allowed range of $N_{c}$ ($0.98(0.94)< N_{c}<2.01(1.95)$). Then,  we  study  the sign of  $\sin \delta$ in the range of $N_{c}$ allowed by experimental data in all these models. We  discuss the model  dependence of our results  in detail.
\newline
The remainder of this paper is organized as follows. In Section 2, we present the form of the effective Hamiltonian and the values of Wilson coefficients. In Section 3,  we give the formalism for the CP violating asymmetry in  $B^{+} \rightarrow \rho^{0}(\omega) \pi^{+} \rightarrow \pi^{+} \pi^{-} \pi^{+}$,  for  all the  models which will be checked. We  also show  numerical results in this section (asymmetry, $a$, the value of  $\sin \delta$). In Section 4, we calculate branching ratios for $B^{+} \rightarrow \rho^{0} \pi^{+} $ and  $B^{0} \rightarrow \rho^{+} \pi^{-} $ and   present numerical results over  the  range of $N_{c}$ allowed  by  the CLEO data. In last section, we  summarize our results  and suggest  further  work.
\section{The effective Hamiltonian}
In order to calculate the direct CP violating aymmetry in  hadronic  decays, one can  use the following effective weak Hamiltonian, based on the operator product expansion~\cite{ref7},
\begin{eqnarray}
{\cal H}_{\bigtriangleup B=1}=\frac {G_{F}}{\sqrt 2} [ \sum_{q=d,s} V_{ub}V_{uq}^{\ast}(c_{1}O_{1}^{u} + c_{2}O_{2}^{u})- V_{tb}V_{tq}^{\ast} \sum_{i=3}^{10} c_{i}O_{i} ] + h.c. ,
\end{eqnarray}
where $c_{i} (i=1,{\cdots},10)$ are the Wilson coefficients. They are calculable in renormalization group improved   pertubation theory and are scale dependent. In the present case, we  use their values at the renormalization scale $\mu \approx  m_{b}$.  The operators $O_{i}$ have the following  form,

\vspace{-3em}
\begin{eqnarray*}
\begin{array}{ll}
O_{1}^{u}=\bar{q}_{\alpha} \gamma_{\mu}(1-\gamma{_5})u_{\beta}\bar{u}_{\beta} \gamma^{\mu}(1-\gamma{_5})b_{\alpha}, \; \; \;
O_{2}^{u}=\bar{q} \gamma_{\mu}(1-\gamma{_5})u\bar{u} \gamma^{\mu}(1-\gamma{_5})b,  \nonumber \; \; \; \; \; \; \; \;  \; \; \;\; \;  \;    \\
\end{array}
\end{eqnarray*}
\vspace{-6em}
\begin{eqnarray}
O_{3}=\bar{q} \gamma_{\mu}(1-\gamma{_5})b \sum_{q\prime}
\bar{q}^{\prime}\gamma^{\mu}(1-\gamma{_5})q^{\prime}, \; \; \; 
O_{4}=\bar{q}_{\alpha} \gamma_{\mu}(1-\gamma{_5})b_{\beta} 
\sum_{q\prime}\bar{q}^{\prime}_{\beta}\gamma^{\mu}(1-\gamma{_5})q^{\prime}_{\alpha},     \nonumber\; \; \; \; & &\\
O_{5}=\bar{q} \gamma_{\mu}(1-\gamma{_5})b \sum_{q'}\bar{q}^
{\prime}\gamma^{\mu}(1+\gamma{_5})q^{\prime}, \; \; \; 
O_{6}=\bar{q}_{\alpha} \gamma_{\mu}(1-\gamma{_5})b_{\beta} 
\sum_{q'}\bar{q}^{\prime}_{\beta}\gamma^{\mu}(1+\gamma{_5})q^{\prime}_{\alpha},    \nonumber\; \; \;\; & & \\
O_{7}=\frac{3}{2}\bar{q} \gamma_{\mu}(1-\gamma{_5})b \sum_{q'}e_{q^{\prime}}
\bar{q}^{\prime} \gamma^{\mu}(1+\gamma{_5})q^{\prime},  \; \; \; \; \; \;\; \; \; \;  \; \;  \; \; \;  \; \; \;\;  \; \; \; \; \; \; \; \; \;\; \; \; \;  \; \; \;\;\; \; \; \;  \; \; \;\; \; \; \;  \; \; \; \; \; \; \; \;\; \;\;\; \; \;  \; & \nonumber \\ 
O_{8}=\frac{3}{2}\bar{q}_{\alpha} \gamma_{\mu}(1-\gamma{_5})b_{\beta} 
\sum_{q'}e_{q^{\prime}}\bar{q}^{\prime}_{\beta}\gamma^{\mu}(1+\gamma{_5})q^{\prime}_{\alpha},\; \; \; \; \; \;\; \; \; \;  \; \;  \; \; \; \; \; \; \;\;  \; \; \; \; \; \; \; \; \;\; \; \; \;  \; \; \;\;\;\; \;\;\;  \;  \; \; \;\;\; \; \; \;  \; \; \; \; \; \; &  \nonumber \\ 
O_{9}=\frac{3}{2}\bar{q} \gamma_{\mu}(1-\gamma{_5})b \sum_{q'}e_{q^{\prime}}
\bar{q}^{\prime} \gamma^{\mu}(1-\gamma{_5})q^{\prime},  \; \; \; \; \; \;\; \; \; \;  \; \;  \; \; \; \;  \; \;\;  \; \; \; \; \; \;\; \; \; \;\; \; \; \;  \; \; \;\;\; \; \; \;  \; \; \;\; \; \; \;  \; \; \; \; \; \; \; \;\; \; \;\; \;  \;  &  \nonumber \\
 O_{10}=\frac{3}{2}\bar{q}_{\alpha} \gamma_{\mu}(1-\gamma{_5})b_{\beta} 
\sum_{q'}e_{q^{\prime}}\bar{q}^{\prime}_{\beta}\gamma^{\mu}(1-\gamma{_5})q^{\prime}_{\alpha},\; \; \; \; \; \;\; \; \; \;  \; \;  \; \; \; \; \; \; \;\;  \; \; \; \; \; \; \; \; \;\; \; \; \;  \; \; \;\;\;\; \;\;\;  \;   \; \;\;\; \; \; \;  \; \; \; \; \; \;  & 
\end{eqnarray}
where $\alpha$ and $\beta$ are color indices, and $q^{\prime}=u,\;d \; {\rm or} \; s$ quarks.
In Eq.(2), $O_{1}$ and $O_{2}$ are the tree level operators,  $O_{3}-O_{6}$ are QCD penguin operators, and $O_{7}-O_{10}$ arise from electroweak penguin diagrams. 

The Wilson coefficients, $c_{i}$, are known to the next-to-leading logarithmic order. At the scale $\mu=m_{b}=5$GeV,  they take  values  the following values~\cite{ref8,ref9}:
\vspace{-1.0em}
\begin{eqnarray*}
\begin{array}{ll}
\vspace{0.5em}
 c_{1}=-0.3125,\;\;\;\; c_{2}=1.1502,   \\
\vspace{0.5em}
 c_{3}=0.0174 ,\;\;\;\; c_{4}=-0.0373,   \\
\vspace{0.5em}
 c_{5}=0.0104 ,\;\;\;\; c_{6}=-0.0459,  \\
\vspace{0.5em}
 c_{7}=-1.050 \times 10^{-5}, \;\;\;\; c_{8}=3.839 \times 10^{-4},  
\end{array}
\end{eqnarray*}
\vspace{-5em}
\begin{eqnarray}
\!\!\!\!\!\!\!\! \!\!\!\!\!\!\!c_{9}=-0.0101, \;\;\;\; 
c_{10}=1.959 \times 10^{-3}.  & &  
\end{eqnarray}
To be consistent, the matrix elements of the operators $O_{i}$ should also be renormalized to the one-loop order. This results in the effective Wilson coefficients, $c_{i}^{\prime}$, which  satisfy the constraint,
\begin{eqnarray}
c_{i}(m_{b})\langle O_{i}(m_{b})\rangle=c_{i}^{\prime}{\langle O_{i}\rangle}^{tree}, 
\end{eqnarray}
where ${\langle O_{i}\rangle}^{tree}$ are the matrix elements at the tree level, which will be evaluated in the factorization approach. From Eq.(4), the relations between $c_{i}^{\prime}$ and $c_{i}$ are~\cite{ref8,ref9},
\begin{eqnarray*}
\begin{array}{ll}
\vspace{0.5em}
c_{1}^{\prime}=c_{1},\;\;\;\; c_{2}^{\prime}=c_{2},  \\
\vspace{0.5em}
 c_{3}^{\prime}=c_{3}-P_{s}/3,\;\;\;\; c_{4}^{\prime}=c_{4}+P_{s},  \\
\vspace{0.5em}
c_{5}^{\prime}=c_{5}-P_{s}/3,\;\;\; c_{6}^{\prime}=c_{6}+P_{s},  \\
\vspace{0.5em}
c_{7}^{\prime}=c_{7}+P_{e},\;\;\;\;\; c_{8}^{\prime}=c_{8},  \\
\end{array}
\end{eqnarray*}
\vspace{-5em}
\begin{eqnarray}
\!\!\!\!\!\!\!\! \!\!\!\!\!\!\!c_{9}^{\prime}=c_{9}+P_{e},\;\;\;\; 
c_{10}^{\prime}=c_{10}, & & 
\end{eqnarray}
\vspace{-2em}
where
\begin{eqnarray*}
\begin{array}{ll}
%
%\vspace{0.5em}
%
P_{s}=(\alpha_{s}/8\pi)c_{2}(10/9+G(m_{c},\mu,q^{2})),  \\
P_{e}=(\alpha_{em}/9\pi)(3c_{1}+c_{2})(10/9+G(m_{c},\mu,q^{2})), \\
\end{array}
\end{eqnarray*}
with
\vspace{-1em}
\begin{eqnarray*}
 G(m_{c},\mu,q^{2})=4\int_{0}^{1}dxx(x-1){\rm ln} \frac{m_{c}^{2}-x(1-x)q^{2}}{\mu^{2}}.
\end{eqnarray*}
Here $q^{2}$ is   the typical  momentum transfer of the gluon or photon in the penguin diagrams.
\vspace{-1em}
\newline
$G(m_{c},\mu,q^{2})$ has the following explicit expression~\cite{ref10}, 
\begin{eqnarray}
\Re e\; G= \frac{2}{3} \left({\rm ln} \frac{m_{c}^{2}}{\mu^{2}}- \frac{5}{3}-4 \frac{m_{c}^{2}}{q^{2}}+\left(1+2\frac{m_{c}^{2}}{q^{2}}\right)\sqrt{1-4\frac{m_{c}^{2}}{q^{2}}}{\rm ln} \frac{1+\sqrt{1-4\frac{m_{c}^{2}}{q^{2}}}}{1-\sqrt{1-4\frac{m_{c}^{2}}{q^{2}}}}\right), & & \nonumber \\
\Im m\; G= -\frac{2}{3}\left(1+2\frac{m_{c}^{2}}{q^{2}}\right)\sqrt{1-4\frac{m_{c}^{2}}{q^{2}}}. \;\;\;\;\;\;\;\;\;\;\;\;\;\;\;\;\;\;\;\;\;\;\;\;\;\;\;\;\;\;\;\;\;\;\;\;\;\;\;\;\;\;\;\;\;\;\;\;\;\;\;\;\;\;\;\;\; & 
\end{eqnarray}
Based on simple arguments  at the quark level, the value of $q^{2}$ is chosen in the range $0.3 < q^{2}/m_{b}^{2} < 0.5$~\cite{ref3,ref4}. From Eqs.(5,6) we can obtain numerical values for   $c_{i}^{\prime}$.
\newline
When $ q^{2}/m_{b}^{2}=0.3$,
\vspace{-2em}
\begin{eqnarray}
c_{1}^{\prime}=-0.3125,\;\;\;\;  c_{2}^{\prime}=1.1502, \;\;\;\;\;\;\;\;\;\;\;
\;\;\;\;\;\;\;\;\;\;\;\;\;\;\;\;\;\;\;\;\;\;\;\;\;\;\;\;\;\;\;\;\;\;\;
\;\;\;\;\;\;\;\;\;\;\;\;\;\;\;\;\;\;\;\;\;\;\;\;\; &  \nonumber \\
c_{3}^{\prime}= 2.433 \times 10^{-2} + 1.543 \times 10^{-3}i, \;\;\;\; c_{4}^{\prime}= -5.808 \times 10^{-2} -4.628 \times 10^{-3}i, \;\;  &  \nonumber \\
c_{5}^{\prime}=1.733 \times 10^{-2}+ 1.543 \times 10^{-3}i,\;
\;\;\; c_{6}^{\prime}=-6.668 \times 10^{-2}- 4.628 \times 10^{-3}i, \;\; &  \nonumber  \\
c_{7}^{\prime}=-1.435 \times 10^{-4} -2.963 \times 10^{-5}i,\;\; \;
\;c_{8}^{\prime}=3.839 \times 10^{-4}, \;\;\;\;\;\; \;\;\;\;\;\;\;\;\;\;\; \;\;\;\;\;\;\;\; \;\;&  \nonumber \\
c_{9}^{\prime}=-1.023 \times 10^{-2} -2.963 \times 10^{-5}i,\;\;
    \;\; c_{10}^{\prime}=1.959 \times 10^{-3}, \;\;\;\;\;\;\;\;\;\;\;\;\;\;\;\;\;\;\;\;\;\;\; \;\;\;  & 
\end{eqnarray}
and when $q^{2}/m_{b}^{2}=0.5$, one has,
\vspace{-1em}
\begin{eqnarray}
 c_{1}^{\prime}=-0.3125,\;\;\;\;  c_{2}^{\prime}=1.1502,  \;\;\;\;\;\;\;\;\;\;\;
\;\;\;\;\;\;\;\;\;\;\;\;\;\;\;\;\;\;\;\;\;\;\;\;\;\;\;\;\;\;\;\;\;\;\;
\;\;\;\;\;\;\;\;\;\;\;\;\;\;\;\;\;\;\;\;\;\;\;\;\; &  \nonumber \\
 c_{3}^{\prime}= 2.120 \times 10^{-2} + 2.174 \times 10^{-3}i,\;\;\;\; c_{4}^{\prime}= -4.869 \times 10^{-2} -1.552 \times 10^{-2}i, \; \; &  \nonumber \\
 c_{5}^{\prime}=1.420 \times 10^{-2} + 5.174 \times 10^{-3}i  ,\;\;\;\; c_{6}^{\prime}=-5.729 \times 10^{-2}- 1.552 \times 10^{-2}i, \;\; &  \nonumber \\
 c_{7}^{\prime}=-8.340 \times 10^{-5} -9.938 \times 10^{-5}i,\;\;\;\;c_{8}^{\prime}=3.839 \times 10^{-4}, \;\;\;\; \;\;\;\;\;\;\;\;\;\;\; \;\;\;\;\;\;\;\;\;\;\;\;&  \nonumber  \\
 c_{9}^{\prime}=-1.017 \times 10^{-2} -9.938 \times 10^{-5}i,\;\;\;\; c_{10}^{\prime}=1.959 \times 10^{-3},\;\;\;\;\;\;\;\;\;\;\;\;\;\;\;\;\;\;\;\;\;\;\;\;\;\;  & 
\end{eqnarray}
\vspace{-1em}
\newline
where we have taken $\alpha_{s}(m_{Z})=0.112, \;\;\;   \alpha_{em}(m_{b})=1/132.2,\;\;\;  m_{b}=5$GeV, and $ \;\; m_{c}=1.35$GeV.
\section{CP violation in $B^{+} \rightarrow  \rho^{0}(\omega)\pi^{+}  \rightarrow  \pi^{+}  \pi^{-} \pi^{+} $}
\subsection{Formalism }
The formalism for CP violation in hadronic B meson  decays is the following. Let $A$ be the amplitude for the decay $B^{+} \rightarrow  \pi^{+}  \pi^{-} \pi^{+}$, then one has
\vspace{-2em}
\begin{eqnarray}
A=\langle  \pi^{+}  \pi^{-} \pi^{+}|H^{T}|B^{+} \rangle + \langle  \pi^{+}  \pi^{-} \pi^{+}|H^{P}|B^{+} \rangle,
\end{eqnarray}
with $H^{T}$ and $H^{P}$ being the Hamiltonians for the tree and penguin operators , respectively. We can define the relative magnitude and phases between these two contributions  as follows,
\begin{eqnarray}
A= \langle  \pi^{+}  \pi^{-}\pi^{+}|H^{T}|B^{+} \rangle [ 1+re^{i\delta}e^{i\phi}],\;\;  & \\
\bar A= \langle \pi^{+}  \pi^{-}  \pi^{-}|H^{T}|B^{-} \rangle [ 1+re^{i\delta}e^{-i\phi}],  & 
\end{eqnarray}
where $\delta$ and $\phi$ are strong and weak phases, respectively. The phase $\phi$ arises from the appropriate combination of CKM matrix elements which is $ \phi={\rm arg}[(V_{tb}V_{td}^{\star})/(V_{ub}V_{ud}^{\star})]$. As a result, $\sin \phi$ is equal to $\sin \alpha$ with $\alpha$ defined in the standard way~\cite{ref11}. The parameter $r$ is the absolute value of the ratio of tree and penguin amplitudes:

\begin{eqnarray}
r \equiv \left| \frac{\langle \rho^{0}(\omega)\pi^{+}|H^{P}|B^{+} \rangle}{\langle\rho^{0}(\omega)\pi^{+}|H^{T}|B^{+} \rangle} \right|.
\end{eqnarray}
The CP violating asymmetry, $a$, can be written as:

\begin{eqnarray}
a \equiv \frac{|A|^{2}-|\bar A|^{2}}{ |A|^{2}+|\bar A|^{2}}=\frac{-2r\sin\delta \sin\phi}{1+2r\cos\delta \cos\phi+r^2}.
\end{eqnarray}
It can be seen explicitly from Eq.(13) that both weak and strong phase differences  are needed to produce CP violation. In order to obtain a  large signal for direct CP violation, we need some mechanism to make both   $\sin\delta$  and  $r$ large. We stress that   $\rho-\omega$ mixing has the dual advantages that the strong phase difference is large (passing through $90^{o}$ at the $\omega$ resonance) and well known~\cite{ref4,ref5}. With this mechanism, to first  order in  isospin violation, we have the following results when the invariant mass of $\pi^{+}\pi^{-}$ is near the $\omega$ resonance mass,

\begin{eqnarray}
\langle \pi^{-}\pi^{+} \pi^{+}|H^{T}|B^{+} \rangle= \frac{g_{\rho}}{s_{\rho}s_{\omega}} \tilde{\Pi}_{\rho \omega}t_{\omega} +\frac{g_{\rho}}{s_{\rho}}t_{\rho}, \\
\langle  \pi^{-}\pi^{+}   \pi^{+}|H^{P}|B^{+} \rangle= \frac{g_{\rho}}{s_{\rho}s_{\omega}} \tilde{\Pi}_{\rho \omega}p_{\omega} +\frac{g_{\rho}}{s_{\rho}}p_{\rho}.
\end{eqnarray}
Here $t_{V} (V=\rho \;{\rm  or} \; \omega) $ is the tree amplitude and $p_{V}$ the penguin amplitude for  producing a vector meson, V, $g_{\rho}$ is the coupling for $\rho^{0} \rightarrow \pi^{+}\pi^{-}$, $\tilde{\Pi}_{\rho \omega}$ is the effective $\rho-\omega$ mixing amplitude, and $s_{V}$  is  from the inverse  propagator of the vector meson V,

\begin{eqnarray}
s_{V}=s-m_{V}^{2}+im_{V}\Gamma_{V}, & & 
\end{eqnarray}
with $\sqrt s$ being the invariant mass of the $\pi^{+}\pi^{-}$ pair.

We stress that the direct coupling $ \omega \rightarrow \pi^{+} \pi^{-} $ is effectively absorbed into $\tilde{\Pi}_{\rho \omega}$~\cite{ref12},  leading  to the explicit $s$ dependence of $\tilde{\Pi}_{\rho \omega}$. Making the expansion  $\tilde{\Pi}_{\rho \omega}(s)=\tilde{\Pi}_{\rho \omega}(m_{\omega}^{2})+(s-m_{w}^{2}) \tilde{\Pi}_{\rho \omega}^{\prime}(m_{\omega}^{2})$, the  $\rho-\omega$ mixing parameters were determined in the fit of Gardner and O'Connell~\cite{ref13}: $\Re e \; \tilde{\Pi}_{\rho \omega}(m_{\omega}^{2})=-3500 \pm 300 {\rm MeV}^{2}, \;\;\; \Im m \; \tilde{\Pi}_{\rho \omega}(m_{\omega}^{2})= -300 \pm 300 {\rm MeV}^{2}$ and  $\tilde{\Pi}_{\rho \omega}^{\prime}(m_{\omega}^{2})=0.03 \pm 0.04$. In practice, the effect of the derivative term is negligible.
%
%\newline
%
From Eqs.(10,14,15) one has,

\begin{eqnarray}
 re^{i \delta} e^{i \phi}= \frac{ \tilde {\Pi}_{\rho \omega}p_{\omega}+s_{\omega}p_{\rho}}{\tilde {\Pi}_{\rho \omega} t_{\omega} + s_{\omega}t_{\rho}}. 
\end{eqnarray}
Defining
\vspace{-2em}
\begin{center}
\begin{eqnarray}
\frac{p_{\omega}}{t_{\rho}} \equiv r^{\prime}e^{i(\delta_{q}+\phi)}, \;\;\;\;
\frac{t_{\omega}}{t_{\rho}} \equiv \alpha e^{i \delta_{\alpha}}, \;\;\;\;
\frac{p_{\rho}}{p_{\omega}} \equiv \beta e^{i \delta_{\beta}}, 
\end{eqnarray}
\end{center}
where $ \delta_{\alpha}, \delta_{\beta}$ and $ \delta_{q}$ are strong phases,  one finds  the following expression from Eq.(18)

\begin{eqnarray}
re^{i\delta}=r^{\prime}e^{i\delta_{q}} \frac{\tilde{\Pi}_{\rho \omega}+ \beta e^{i \delta_{\beta}} s_{\omega}}{s_{\omega}+\tilde{\Pi}_{\rho \omega} \alpha e^{i \delta_{\alpha}}}.
\end{eqnarray}
It will be shown that in the factorization approach, we have $\alpha e^{i \delta_{\alpha}}=1$ in our case. 
Letting
\vspace{-2em}
\begin{eqnarray}
& \beta e^{i \delta_{\beta}}= b+ci, \;\;\;r^{\prime}e^{i\delta_{q}}=d+ei,
\end{eqnarray}
and using Eq.(20), we obtain the following result when $\sqrt s \sim m_{\omega}$
\vspace{-1em}
\begin{eqnarray}
re^{i \delta}= \frac{C+Di}{(s-m_{\omega}^{2}+ \Re e \; \tilde{\Pi}_{\rho \omega})^{2}+ (\Im m \; \tilde{\Pi}_{\rho \omega} +m_{\omega} \Gamma_{\omega})^{2}},
\end{eqnarray}
where
\vspace{-2em}
\begin{eqnarray}
C=(s-m_{\omega}^{2}+ \Re e\; \tilde {\Pi}_{\rho \omega}) \bigg\{ d[ \Re e \;\tilde {\Pi}_{ \rho \omega} +b(s-m_{ \omega}^{2})-cm_{ \omega} \Gamma_{ \omega}] & \nonumber \\ 
  -e [ \Im m \;\tilde { \Pi}_{ \rho \omega} +bm_{ \omega} \Gamma_{ \omega}+c(s-m_{ \omega}^{2})] \bigg\} \;\;\;\;\;\;\;\;\;\;\;\;\;\;\;\;\;\;\;\;\;\;\;\;\;\;\; & \nonumber  \\
   + ( \Im m\; \tilde { \Pi}_{ \rho \omega}  +m_{ \omega} \Gamma_{ \omega}) \bigg\{ e[ \Re e \;\tilde {\Pi}_{\rho \omega} +b(s-m_{ \omega}^{2})-cm_{ \omega} \Gamma_{ \omega}]  & \nonumber \\
  +d[ \Im m \;\tilde { \Pi}_{ \rho \omega} 
+bm_{ \omega} \Gamma_{ \omega}+c(s-m_{ \omega}^{2})] \bigg\}, \;\;\;\;\;\;\;\;\;\;\;\;\;\;\;\;\;\;\;\;\;\;\;\;\;\; & \nonumber \\
 D=(s-m_{\omega}^{2}+ \Re e \;\tilde {\Pi}_{\rho \omega}) \bigg\{ e[ \Re e \;\tilde {\Pi}_{ \rho \omega} +d(s-m_{ \omega}^{2})-cm_{ \omega} \Gamma_{ \omega}] &  \nonumber \\ 
  +d [ \Im m\; \tilde { \Pi}_{ \rho \omega} +bm_{ \omega} \Gamma_{ \omega}+c(s-m_{ \omega}^{2})]  \bigg\} \;\;\;\;\;\;\;\;\;\;\;\;\;\;\;\;\;\;\;\;\;\;\;\;\;\;\;  & \nonumber \\
   - ( \Im m \;\tilde { \Pi}_{ \rho \omega}  +m_{ \omega} \Gamma_{ \omega}) \bigg\{ d[ \Re e \;\tilde {\Pi}_{\rho \omega} +b(s-m_{ \omega}^{2})-cm_{ \omega} \Gamma_{ \omega}]  & \nonumber \\
  -e[ \Im m\; \tilde { \Pi}_{ \rho \omega} 
+bm_{ \omega} \Gamma_{ \omega}+c(s-m_{ \omega}^{2})] \bigg\}. \;\;\;\;\;\;\;\;\;\;\;\;\;\;\;\;\;\;\;\;\;\;\;\;\;\; & 
\end{eqnarray}
$\beta e^{i \delta_{\beta}}$ and $r^{\prime}e^{i \delta_{q}}$ will be calculated later. Then, from Eq.(22) we can obtain $r\sin \delta$ and $r\cos\delta$. In order to get the CP violating asymmetry, $a$, in Eq.(13), $\sin\phi$ and $\cos\phi$ are needed, where $\phi$ is determined by the CKM matrix elements. In the Wolfenstein parametrization~\cite{ref14}, one has,

\begin{eqnarray}
\sin\phi= \frac{\eta}{\sqrt {[\rho(1-\rho)-\eta^{2}]^{2}+\eta^{2}}}, \\
\cos\phi= \frac{\rho(1-\rho)-\eta^{2}}{\sqrt {[\rho(1-\rho)-\eta^{2}]^{2}+\eta^{2}}}.
\end{eqnarray}
\vspace{1em}
\subsection{Calculational Details}
 With the Hamiltonian given in Eq.(1), we are ready to evaluate the matrix elements for $B^{+}\rightarrow   \rho^{0}(\omega)\pi^{+}$.  In the factorization approximation, either the $\rho^{0}(\omega)$ or the  $\pi^{+}$ is generated by one current which has the appropriate  quantum numbers in the Hamiltonian.
For this decay process, two kinds of matrix element products are involved after factorization; schematically (i.e. omitting Dirac matrices and color labels)  $\langle \rho^{0}(\omega)|(\bar{u}u)|0\rangle \langle\pi^{+}|(\bar{d}b)|B^{+}\rangle $ and $ \langle\pi^{+}|(\bar{d}u)|0\rangle \langle\rho^{0}(\omega)|(\bar{u}b)|B^{+}\rangle$. We will calculate them in some phenomenological quark models. 
\newline
The matrix elements for $B \rightarrow X$ and  $B \rightarrow X^{\star}$ (where X and $ X^{\star}$ denote pseudoscalar and vector mesons, respectively) can be decomposed as~\cite{ref15},
\vspace{-1em}
\begin{eqnarray}
\langle X|J_{\mu}|B \rangle =\left( p_{B} + p_{X}- \frac{m_{B}^{2}-m_{X}^{2}}{k^{2}}k \right)_{\mu} F_{1}(k^{2})+\frac{m_{B}^{2}-m_{X}^{2}}{k^{2}}k_{\mu}F_{0}(k^{2}), \;\;\;\;\;\;&
\end{eqnarray}
\vspace{-5em}
\begin{eqnarray}
\langle X^{\star}|J_{\mu}|B \rangle=\frac{2}{m_{B}+m_{X^{\star}}} \epsilon_{\mu \nu \rho \sigma} \epsilon^{\star \nu} p_{B}^{\rho} p_{X^{\star}}^{\sigma}  V(k^{2}) +i \{ \epsilon_{\mu}^{\star}(m_{B}+m_{X^{\star}})A_{1}(k^{2}) & \nonumber \\
- \frac{\epsilon^{\star}  \cdot k}{m_{B}+m_{X^{\star}}} (P_{B}+P_{X^{\star}})_{\mu}A_{2}(k^{2})- \frac {\epsilon^{\star}  \cdot k}{k^{2}}2m_{X^{\star}} \cdot k_{\mu}A_{3}(k^{2})
\} \;\; & \nonumber \\
+i \frac{\epsilon^{\star} \cdot k}{ k^{2}}2m_{X^{\star}} \cdot k_{\mu}A_{0}(k^{2}),\;\;\;\;\;\;\;\;\;\;\;\;\;\;\;\;\;\;\;\;\;\;\;\;\;\;\;\;\;\;\;\;\;\;\;\;\;\;\;\;\;\;\;\;\;\;\;\;\;\;\;\;
\end{eqnarray}
%
%\vspace{1em}
where $J_{\mu}$  is the weak current ($J_{\mu}=\bar{q}\gamma^{\mu}(1-\gamma_{5})b \;\;       {\rm with} \;\; q=u,d$),  $k=p_{B}-p_{X(X^{\star})}$ and $\epsilon_{\mu}$ is the polarization vector of $X^{\star}$. The form factors included in our calculations  satisfy: $F_{1}(0)=F_{0}(0), \;\;\;  A_{3}(0)=A_{0}(0)$ and  
$A_{3}(k^{2})= \frac {m_{B}+m_{X^{\star}}}{2m_{X^{\star}}}A_{1}(k^{2})- \frac {m_{B}-m_{X^{\star}}}{2m_{X^{\star}}}A_{2}(k^{2}).$ 
%
%
%\vspace{1em}
Using the decomposition in Eqs.(25, 26), one has,
%\vspace{-1em}
%
\begin{eqnarray}
t_{\rho}=  m_{B}|\vec{p}_{\rho}|  \left[ (c_{1}^{\prime}+\frac {1}{N_{c}}c_{2}^{\prime})  f_{\rho}F_{1}(m_{\rho}^{2})+ (c_{2}^{\prime}+\frac {1}{N_{c}}c_{1}^{\prime} )    f_{\pi}A_{0}(m_{\pi}^{2}) \right],
\end{eqnarray}
%
%\vspace{1em}
where $f_{\rho}$ and $f_{\pi}$ are the decay constants of the $\rho$ and $\pi$, respectively, and $\vec{p}_{\rho}$  is the three momentum of the $\rho$.
In the same way, we find $t_{\omega}=t_{\rho}$, so that
\begin{eqnarray}
\alpha e^{i \delta_{\alpha}}=1. 
\end{eqnarray}
\vspace{-1em}
After calculating the penguin operator contributions, one has,
%
%\vspace{1em}
\begin{eqnarray}
\beta e^{i \delta_{\beta}}=
\frac{ m_{B}| \vec{p}_{\rho}|}{p_{\omega}} \Bigg\{ (c_{4}^{\prime}+\frac{1}{N_{c}}c_{3}^{\prime})[-f_{\rho}F_{1}(m_{\rho}^{2})+f_{\pi}A_{0}(m_{\pi}^{2})] \;\;\;\;\;\;\;\;\;\;\;\;\;\;\;\;\;\;\; \;\;\;\;\;\;\;\;\;\;\;\;\;\;\;\;\;\;\;\;\;\;\;\;\;\;\; & \nonumber \\
+\frac{3}{2}[(c_{7}^{\prime}+\frac{1}{N_{c}}c_{8}^{\prime})+(c_{9}^{\prime}+\frac{1}{N_{c}}c_{10}^{\prime})]f_{\rho}F_{1}(m
_{\rho}^{2})  \;\;\;\;\;\;\;\;\;\;\;\;\;\;\;\;\;\;\;\;\;\;\;\;\;\;\;\;\;\;\;\;\;\;\;\;\;\;\;\;\; \; & \nonumber \\
-[(c_{6}^{\prime}+\frac{1}{N_{c}}c_{5}^{\prime})+(c_{8}^{\prime}+\frac{1}{N_{c}}c_{7}^{\prime})] \left[ \frac{2m_{\pi}^{2}f_{\pi}A_{0}(m_{\pi}^{2})}{(m_{u}+m_{d})(m_{b}+m_{u})}\right]   \;\;\;\;\;\;\;\;\;\;\;\;\;\;\;\;\;\;\;\;\;\;\;& \nonumber \\
+ (c_{10}^{\prime}+\frac{1}{N_{c}}c_{9}^{\prime})[\frac{1}{2}f_{\rho}F_{1}(m_{\rho}^{2})+f_{\pi}A_{0}(m_{\pi}^{2})]\Bigg\},\;\;\;\;\;\;\;\;\;\;\;\;\;\;\;\;\;\;\;\;\;\;\;\;\;\;\;\;\;\;\;\;\;\;\;\;\;\;\;\;\; & \nonumber 
\end{eqnarray}
\vspace{-4em}
\begin{eqnarray}
r^{\prime}e^{i \delta_{q}}=-
\frac{p_{\omega}}{(c_{1}^{\prime}+\frac {1}{N_{c}}c_{2}^{\prime})  f_{\rho}F_{1}(m_{\rho}^{2})+ (c_{2}^{\prime}+\frac {1}{N_{c}}c_{1}^{\prime} )    f_{\pi}A_{0}(m_{\pi}^{2})}\left| \frac{V_{tb}V_{td}^{\star}}{V_{ub}V_{ud}^{\star}}\right|, \;\;\;\;\; \;\;\;\;\;\;\;\; \;\;\;\;\;\;\;\;\;\;\;\;\;\; & & 
\end{eqnarray}
%
%\vspace{1em}
where
\begin{eqnarray*}
p_{\omega}= m_{B}|\vec{p}_{\rho}| \Bigg\{ 2 \left[ (c_{3}^{\prime}+\frac {1}{N_{c}}c_{4}^{\prime})+(c_{5}^{\prime}+\frac {1}{N_{c}}c_{6}^{\prime})\right]f_{\rho}F_{1}(m_{\rho}^{2})  \;\;\;\;\; \;\;\;\;\;\;\;\;\;\;\;\;\;\;\;\;\;\;\;\;\;\;\;\;\;\;\;\; \;\;\;\;\;\;\;\;\;\;\;\;\;\;\;\;\;\; \;\;\;\;\;\;\;\;\;\;\;\;\;\;\;\;\;\;\;\;\;\;\;\;\; & & \nonumber  \\
+  \frac{1}{2}\left[(c_{7}^{\prime}+\frac {1}{N_{c}}c_{8}^{\prime})+(c_{9}^{\prime}+\frac {1}{N_{c}}c_{10}^{\prime})\right] f_{\rho}F_{1}(m_{\rho}^{2})  \;\; \;\;\;\;\;\;\;\;\;\;\;\;\;\;\;\;\;\;\;\;\;\;\;\;\;\;\;\;\;\;\;\;\;\;\;\;\;\;\;\;\;\;\;\;\;\;\;\;\;\;\;\;\; \;\;\;\;\;\;\;\;\;\;\;\;\;\;\;\;\;\; & & \nonumber \\
- 2 \left[(c_{8}^{\prime}+\frac {1}{N_{c}}c_{7}^{\prime})+(c_{6}^{\prime}+\frac {1}{N_{c}}c_{5}^{\prime}) \right] \left[ \frac{m_{\pi}^{2}f_{\pi}A_{0}(m_{\pi}^{2})}{(m_{u}+m_{d})(m_{b}+m_{u})}\right]  \;\;\;\;\;  \;\;\;\;\;\; \;\;\;\;\;\;\;\;\;\; \;\;\;\;\;\; \;\;\;\;\; \;\;\;\;\;\; \;\;\;\;\; \;\;\;\;\;\; \;\;\; & \nonumber \\ 
+  (c_{4}^{\prime}+\frac {1}{N_{c}}c_{3}^{\prime}) \left[ f_{\pi}A_{0}(m_{\pi}^{2})+f_{\rho}F_{1}(m_{\rho}^{2})\right]  \;\;\;\;\;\; \;\;\;\;\;\;\;\;\;\;\;\; \;\;\;\;\;\;\;\;\;\;\;\; \;\;\;\;\;\;\;\;\;\;\;\;\;\;\;\;\;\; \;\;\;\;\;\;\;\;\;\;\;\;\;\;\;\;\;\; \;\;\;\;\;\;\;\; \;\;\;\;\;\;  & &   \nonumber \\
+   (c_{10}^{\prime}+\frac {1}{N_{c}}c_{9}^{\prime}) \left[ f_{\pi}A_{0}(m_{\pi}^{2})-\frac{1}{2}f_{\rho}F_{1}(m_{\rho}^{2})\right] \Bigg\},  \;\;\;\; \;\;\;\; \;\;\;\;  \;\;\;\;\;\;\;\;\;\; \;\;\;\; \;\;\;\;\;\;\;\; \;\;\;\; \;\;\;\;  \;\;\;\;\;\; \;\;\;\;\;\;\;\; \;\;\;\; \;\;\;\; \;\;\;\;  \;\;\;\;  & &   
\end{eqnarray*}
and
%\vspace{1em}
%
\begin{eqnarray}
\left| \frac{V_{tb}V_{td}^{\star}}{V_{ub}V_{ud}^{\star}}\right|
=\frac{\sqrt{{(1-\rho)}^{2}+{\eta}^{2}}}
{(1- {\lambda}^{2}/2)\sqrt{{\rho}^{2}+{\eta}^{2}}}=\left( 1- \frac{\lambda^{2}}{2} \right)^{-1} \left| \frac{\sin \gamma}{\sin \beta} \right|. & 
\end{eqnarray}
\subsection{Numerical Results}
In our numerical calculations  we have several parameters: $q^{2}, N_{c}$ and the CKM matrix elements in the Wolfenstein parametrization. As mentioned in Section 2, the value of $q^{2}$ is conventionally chosen to be in the range $0.3<q^{2}/{m_{b}}^{2}<0.5$. The CKM matrix, which should be determined from experimental data,  has the following form  in term of the Wolfenstein parameters, $ A,\; \lambda,\; \rho, \; \eta $~\cite{ref14}:
\begin{eqnarray}
V= \left( \begin{array}{ccc}
1-\frac{1}{2} \lambda^{2} &  \lambda                    & A\lambda^{3}(\rho-i\eta) \\
-\lambda                  & 1-\frac{1}{2}\lambda^{2}    & A\lambda^{2}             \\
A\lambda^{3}(1-\rho-i\eta)& -A\lambda^{2}               &      1                   \\
\end{array}  \right),
\end{eqnarray}
\vspace{-1em}
where $O(\lambda^{4})$ corrections are neglected.
We use $\lambda=0.2205$, $A=0.815$ and the range  for $\rho$ and $\eta$  as the following ~\cite{ref16,ref17},
\begin{eqnarray}
0.09 < \rho < 0.254, \;\;\;\;  \;\;\;\;\; 0.323 < \eta <0.442.
\end{eqnarray}
 The form factors  $F_{1}(m_{\rho}^{2})$ and $A_{0}(m_{\pi}^{2})$  depend on the inner structure of the  hadrons. Under the nearest pole dominance assumption, the $k^{2}$ dependence of the form factors is:

for model 1(2)~\cite{ref15,ref18}:
\begin{eqnarray}
F_{1}(k^{2})=\frac{h_{1}}{1-\frac{k^{2}}{m_{1}^{2}}}, \;\;\;  \;\;\; 
A_{0}(k^{2})=\frac{h_{A_{0}}}{1-\frac{k^{2}}{m_{A_{0}}^{2}}},
\end{eqnarray} 
where $h_{1}=0.330(0.625), \;\;\;\; h_{A_{0}}=0.28(0.34), \;\;\;\; m_{1}=5.32{\rm GeV}, \;\;\;\; m_{A_{0}}=5.27{\rm GeV}, $

for model 3(4)~\cite{ref15,ref18,ref19}:
\begin{eqnarray}
F_{1}(k^{2})=\frac{h_{1}}{\left( 1-\frac{k^{2}}{m_{1}^{2}} \right)^{2}}, \;\;\;  \;\;\; 
A_{0}(k^{2})=\frac{h_{A_{0}}}{\left( 1-\frac{k^{2}}{m_{A_{0}}^{2}}\right)^{2}},
\end{eqnarray} 
where $h_{1}=0.330(0.625), \;\;\;\; h_{A_{0}}=0.28(0.34), \;\;\;\; m_{1}=5.32{\rm GeV}, \;\;\;\; m_{A_{0}}=5.27{\rm GeV},$

for model 5~\cite{ref20,ref21}:
\begin{eqnarray}
F_{1}(k^{2})=\frac{h_{1}}{1-a_{1}\frac{k^{2}}{m_{B}^{2}}+b_{1}\left( \frac{k^{2}}{m_{B}^{2}}\right)^{2}}, \;\;\;  \;\;\; 
A_{0}(k^{2})=\frac{h_{A_{0}}}{1-a_{0}\frac{k^{2}}{m_{B}^{2}}+b_{0}\left( \frac{k^{2}}{m_{B}^{2}}\right)^{2}},
\end{eqnarray} 
where $h_{1}=0.305,     \;\;\;\; h_{A_{0}}=0.372,    \;\;\; $
 $ a_{1}=0.266,   \;\;\;\;   b_{1}=-0.752,  \;\;\;\;  a_{0}=1.4,   \;\;\;\;     b_{0}=0.437$.    
\newline
The  decay  constants used  in our calculations are: 
$  f_{\rho}=f_{\omega}=221{\rm MeV} \; {\rm and} \; f_{\pi}=130.7{\rm MeV}$. 
\newline
In the numerical calculations, it is found that for a fixed $N_{c}$, there is a maximum value, $a_{max}$,  for the CP violating parameter, $a$,  when the invariant mass of the $\pi^{+}\pi^{-}$ is in the vicinity of the $\omega$ resonance. The results are  shown in Figs.1 and 2, for $k^{2}/m_{b}^{2}=0.3(0.5)$ and $N_{c}$ in the range $0.98(0.94)<N_{c}<2.01(1.95)$ -- for reasons  which will be explained later   (Section 4). We investigate five models with different  form factors to study the model dependence of  $a$. It appears that this dependence is strong (Table 1).

The maximum asymmetry parameter, $a_{max}$, varies from  $-24\%(-19\%)$ to $-59\%(-48\%)$ for $N_{c}$ in the chosen range,  $k^{2}/m_{b}^{2}=0.3(0.5)$ and the range of CKM matrix elements indicated earlier. If we look at the numerical results for the  asymmetries (Table 1) for $N_{cmax}=2.01(1.95)$ and $k^{2}/m_{b}^{2}=0.3(0.5)$, we obtain for  models  1, 3, 5 an asymmetry, $a_{max}$, around $-27.3\%(-21.6\%)$ for the set $(\rho_{max},\eta_{max})$, and around $-44.3\%(-35.0\%)$ for the set $(\rho_{min},\eta_{min})$. We find a ratio equal to $1.62(1.62)$ between the asymmetries associated with the  upper and lower limits of ($\rho,\eta$). The reason why the maximum asymmetry, $a_{max}$,  can have large variation,  comes from   the  $b \rightarrow d$ transition, where $V_{td}$ and $V_{ub}$ appear. These  are functions  of  ($\rho,\eta$) and   contribute to the asymmetry (Eq.31) through  the ratio between the $\omega$ penguin diagram and the $\rho$ tree diagram.

For  models 2 and 4,  one has a  maximum  asymmetry, $a_{max}$,  around $-37\%(-28\%)$ for the set  $(\rho_{max},\eta_{max})$ and around  $-59\%(-46\%)$ for the set  $(\rho_{min},\eta_{min})$. We  find a ratio  between the asymmetries equal to $1.59(1.64)$  in  this case. The difference between these two sets of  models comes from the magnitudes  of the form factors, where $F_{1}(k^{2})$ is  larger  for models 2 and 4 than for  models 1, 3 and 5. Now, if we look at the  numerical results for the  asymmetry for $N_{cmin}=0.98(0.94)$, we find, for models 1, 3, 5,   $k^{2}/m_{b}^{2}=0.3(0.5)$,  and the set $(\rho_{max},\eta_{max})$,  an asymmetry, $a_{max}$,  around  $-31.3\%(-25.6\%)$, and for   the set  $(\rho_{min},\eta_{min})$ we find an asymmetry, $a_{max}$, around  $-50.3\%(-42.0\%)$. In this case, one has a ratio equal to 1.61(1.64). Finally, for models 2 and 4, we get  $-36\%(-29\%)$  for  the set $(\rho_{max},\eta_{max})$  and  $-57\%(-48\%)$ for the set  $(\rho_{min},\eta_{min})$ with a ratio equal to 1.58(1.65).

These results show explicitly the dependence of the CP violating asymmetry  on form factors, CKM matrix elements and the effective parameter $N_{c}$. For the CKM matrix elements, it appears that if we take their  upper limit, we obtain a smaller  asymmetry, $a$, and viceversa. The difference between $k^{2}/m_{b}^{2}=0.3(0.5)$ in our results comes from the  renormalization of the matrix elements of the operators in the weak Hamiltonian. Finally, the dependence on  $N_{c}$ comes from  the fact that $N_{c}$  is related to hadronization effects, and consequently  we cannot determine  $N_{c}$  exactly  in our calculations. Therefore,  we treat $N_{c}$ as a free effective parameter. As regards  the ratio between the asymmetries, we have found a ratio equal to 1.61(1.63). This is mainly determined   by the ratio $\sin\gamma / \sin\beta$, and more precisely by $\eta$. In Table 2, we show  the values for the angles $\alpha$, $\beta$, $\gamma$. From all these numerical results, we can conclude that we need to determine  the value of  $N_{c}$ and  the hadronic decay form factors more precisely,  if we want to use the  asymmetry, $a$,   to constrain the CKM matrix elements.

In spite of the uncertainties just discussed, it is vital to realize that the effect of $\rho-\omega$ mixing in the $B \rightarrow \rho \pi$ decay is to remove any ambiguity concerning the strong phase, $\sin \delta$. As the internal top quark dominates the $b \rightarrow d$ transition, the weak phase in the rate asymmetry is proportional to $\sin \alpha \; (=\sin \phi)$, where $\alpha={\rm arg}\left[ - \frac{V_{td}V_{tb}^{\star}}{V_{ud}V_{ub}^{\star}}\right]$,  and  knowing the sign of $\sin \delta$ enables us to determine  that  of $\sin \alpha$ from a measurement of the asymmetry, $a$. We show in Fig.3 that the sign of $\sin \delta$ is always positive in our range, $0.98(0.94)<N_{c}<2.01(1.95)$ for all the   models studied. Indeed, at the $\pi^{+}\pi^{-}$ invariant mass where the asymmetry parameter, $a$,  reaches a maximum, the value of $\sin\delta$ is equal to one  -- provided $\rho-\omega$  mixing is included --  over the entire range of $N_{c}$ and for all the form factors studied. So, we can remove, with the help of asymmetry, $a$,  the uncertainty ${\rm mod}(\pi)$ which appears in $\alpha$ from the usual indirect measurements~\cite{ref5}  which yield  $\sin 2\alpha$. By contrast, in the case where  we do not take $\rho-\omega$ mixing into account, we find a small value for $\sin \delta$. In Figs.3 and 4  we plot the role of $\rho - \omega$ mixing in our calculations.  We  stress that, even though  one has a large  value of $\sin \delta$  around $N_{c}=1$ with no $\rho - \omega$ mixing, one still has a very small value for $r$ (Fig.4), and hence  the CP violating asymmetry, $a$,  remains very small in that case.
\section{Branching ratios for $B^{+} \rightarrow \rho^{0}  \pi^{+}$ and $B^{0} \rightarrow \rho^{+} \pi^{-} $ }
\subsection{Formalism }
With the factorized decay amplitudes, we can compute the decay rates using  by the following expression~\cite{ref19},
\begin{eqnarray}
\Gamma(B \rightarrow VP)=\frac{\vec{|p_{\rho}|}^{3}}{8\pi m_{V}^{2}}
\left|\frac{A(B \rightarrow VP)}{\epsilon \cdot p_{B}}\right|^{2}, 
\end{eqnarray}
where
\begin{eqnarray}
|\vec{p_{\rho}}|=\frac{ \sqrt{ [m_{B}^{2}-(m_{1}+m_{2})^{2}][m_{B}^{2}-(m_{1}-m_{2})^{2}]}}{2m_{B}} 
\end{eqnarray}
is the c.m. momentum of the decay particles, $m_{1} (m_{2})$ is the mass of the vector (pseudoscalar) V(P),
and  $A(B \rightarrow VP)$ is the decay amplitude: 
\begin{eqnarray}
A(B \rightarrow VP)=\frac{G_{F}}{\sqrt{2}} \sum_{i=1,10}V_{u}^{T,P}a_{i}\langle VP | O_{i}| B \rangle.
\end{eqnarray}
Here  $V_{u}^{T,P}$ is CKM factor: 
\begin{eqnarray*}
V_{u}^{T}=|V_{ub}V_{ud}^{\star}| \;\;\;\; \mbox{for} \;\;\;\; i=1,2 \;\;\;\; \mbox{and} & V_{u}^{P}=|V_{tb}V_{td}^{\star}| \;\;\;\; \mbox{for} \;\;\;\; i=3, \cdots, 10
\end{eqnarray*}
where the effective parameters  are the following  combinations 
\begin{eqnarray*}
a_{2j}=c_{2j}^{\prime}+\frac{1}{N_{c}}c_{2j-1}^{\prime},   \;\;\; a_{2j-1}=c_{2j-1}^{\prime}+\frac{1}{N_{c}}c_{2j}^{\prime}, \;\;  {\rm for} \;\; j=1, \cdots, 5
\end{eqnarray*}
and $\langle VP | O_{i}| B \rangle$ is a matrix element  which is evaluated in the factorization approach. In the Quark Model, the diagram coming from the $B^{+} \rightarrow \rho^{0} \pi^{+}$ decay is the only one contribution. In our case, to be consistent, we should also take into account the  $\rho - \omega$ mixing contribution when we calculate the branching ratio since we are working to the first order of  isospin violation.
Explicitly, we obtain for $B^{+} \rightarrow \rho^{0} \pi^{+}$,
\begin{eqnarray}
BR(B^{+} \rightarrow \rho^{0} \pi^{+})=\frac{G_{F}^{2}|\vec{p}_{\rho}|^{3}}{32 \pi \Gamma_{B^{+}}}\Bigg|\bigg[V_{u}^{T}A^{T}_{\rho^{0}}(a_{1},a_{2})-V_{u}^{P}A^{P}_{\rho^{0}}(a_{3}, \cdots, a_{10})\bigg]   \nonumber \;\;\;\;\;\;\;\;\;\;\;\;\;\;\;\;\;\;\;\;\;\;\;\;\;\;\;\;\;\;\;\;\;\;\;  & & \\
+\bigg[V_{u}^{T}A^{T}_{\omega}(a_{1},a_{2})-V_{u}^{P}A^{P}_{\omega}(a_{3}, \cdots, a_{10})\bigg]\frac{\tilde{\Pi}_{\rho \omega}}{(s_{\rho}-m_{\omega}^{2})+im_{\omega}\Gamma_{\omega}}\Bigg|^{2},  \;\;\;\;\;\;\;\;\;\;\;\;\;\;\; & &
\end{eqnarray}
where the tree and penguin amplitudes are:
\begin{eqnarray*}
\sqrt{2}A^{T}_{\rho^{0}}(a_{1},a_{2})=a_{1}f_{\rho}F_{1}(m_{\rho}^{2})+a_{2}f_{\pi}A_{0}(m_{\pi}^{2}),\;\;\;\;\;\;\;\;\;\;\;\;\;\;\;\;\;\;\;\;\;\;\;\;\;\;\;\;\;\;\;\;\;\;\;\;\;\;\;\;\;\;\;\;\;\;\;\;\;\;\;\;\;\;\;\;\;\;\;\;\;\;\;\;\;\;\;\;\;\;\;\;\;\;\;\;\;\;\;\;\;\;\;\;\;\;\;\;\;\;\;\;\;\;\;\;\;\;\;\;\;\; & \\
\sqrt{2}A^{P}_{\rho^{0}}(a_{3},  \cdots,   a_{10})=a_{4}\left[ -f_{\rho}F_{1}(m_{\rho}^{2})+f_{\pi}A_{0}(m_{\pi}^{2}) \right] +a_{10}\left[ \frac{1}{2}f_{\rho}F_{1}(m_{\rho}^{2})+f_{\pi}A_{0}(m_{\pi}^{2}) \right] \;\;\;\;\;\;\;\;\;\;\;\;\;\;\;\;\;\;\;\;\;\;\;\;\;\;\;\;\;\;\;\;\;\;\;\;\;\;\;\;\;\;\;\; & \\   +\frac{3}{2}(a_{7}+a_{9})f_{\rho}F_{1}(m_{\rho}^{2})\
-2(a_{6}+a_{8}) \left[ \frac{m_{\pi}^{2}f_{\pi}A_{0}(m_{\pi}^{2})}{(m_{u}+m_{d})(m_{b}+m_{u})}\right],\;\;\;\;\;\;\;\;\;\;\;\;\;\;\;\;\;\;\;\;\;\;\;\;\;\;\;\;\;\;\;\;\;\;\;\; \;\;\;\;\;\;\;\;\; & \\
\sqrt{2}A^{T}_{\omega}(a_{1},a_{2})=a_{1}f_{\rho}F_{1}(m_{\rho}^{2})+a_{2}f_{\pi}A_{0}(m_{\pi}^{2}),\;\;\;\;\;\;\;\;\;\;\;\;\;\;\;\;\;\;\;\;\;\;\;\;\;\;\;\;\; \;\;\;\;\;\;\;\;\;\;\;\;\;\;\;\;\;\;\;\;\;\;\;\;\;\;\;\;\;\;\; \;\;\;\;\;\;\;\; \;\;\;\;\;\;\;\;\;\;\;\;\;\;\;\;\;\;\;\;\;\;\;\;\;\;\;\;\;\;\;\;\;\; & \\
\sqrt{2}A^{P}_{\omega}(a_{3},  \cdots,   a_{10})= \left[ 2(a_{3}+a_{5})+\frac{1}{2}(a_{7}+a_{9})\right]f_{\rho}F_{1}(m_{\rho}^{2})\;\;\;\; \;\;\;\;\;\;\;\;\;\;\;\;\;\;\;\;\;\;\;\;  \;\;\;\;\;\;\;\;\;\;\;\;\;\;\;\;\;\;\; \;\;\;\;\;\;\;\;\;\;\;\;\;\;\;\;\;\;\;\;\;\;\;\;\;\;\;\;\;\;\;\;\;\;\;\;\;\; &  \nonumber  \\
- 2 (a_{8}+a_{6}) \left[ \frac{m_{\pi}^{2}f_{\pi}A_{0}(m_{\pi}^{2})}{(m_{u}+m_{d})(m_{b}+m_{u})}\right]  \;\;\;\;\;\;\;\;\;\;\;\;\;\;\;\;\;\;\;\;\;\;\;\;\;\;\; \;\;\;\;\;\;\;\;\;\;\;\;\;\;\;\;\;\;\;\;\;\;\; \;\;\;\;\;\;\;\;\;\;\;\;\;\;\;\;\;\;\;\;\;\;\;\;\;\;\;\;\;\;\;\;  & \nonumber \\ 
+  a_{4}\left[ f_{\pi}A_{0}(m_{\pi}^{2})+f_{\rho}F_{1}(m_{\rho}^{2})\right]
+   a_{10}\left[ f_{\pi}A_{0}(m_{\pi}^{2})-\frac{1}{2}f_{\rho}F_{1}(m_{\rho}^{2})\right],  \;\;\;\;\;\;\;\;\;\;\;\;\;\;\;\;\;\;\; \;\;\;\;\;\;\;\;\;\;\;\;\;\;\;\;\;\;\; \;\;\;\;\;\ & 
\end{eqnarray*}
where $\langle  \rho^{0}| \bar{u}u | 0 \rangle = \frac{1}{\sqrt{2}}f_{\rho} m_{\rho} \epsilon_{\rho}$ and  $\langle \pi^{+} | \bar{u}d | 0 \rangle = if_{\pi} p_{\mu}$.
\vspace{0.5em}
\newline
For $B^{0} \rightarrow \rho^{+} \pi^{-}$  we obtain,
\begin{eqnarray}
BR(B^{0} \rightarrow \rho^{+} \pi^{-})=\frac{G_{F}^{2}|\vec{p}_{\rho}|^{3}}{16 \pi \Gamma_{B^{0}}}\left|V_{u}^{T}A^{T}_{\rho^{+}}(a_{2})-V_{u}^{P}A^{P}_{\rho^{+}}(a_{3}, \cdots, a_{10})\right|^{2}, \;\;\;\;\;\;\;\;\;\;\;\;\;\;\;\;\;\;\;\;\;\;\;\;\;\;\;\;\;\;\;\;\;\;\;\;  & 
\end{eqnarray}
\vspace{-1em}
where
\vspace{-1em}
\begin{eqnarray*}
A^{T}_{\rho^{+}}(a_{2})=a_{2}f_{\rho}F_{1}(m_{\rho}^{2}),\;\;\;\;\;\;\;\;\;\;\;\;\;\;\;\;\;\;\;\;\;\;\;\;\;\;\; \;\;\;\;\;\;\;\;\;\;\;\;\;\;\;\;\;\;\;\;\;\;\;\;\;\;\;\;\;\;\;\; \;\;\;\;\;\;\;\;\;\;\;\;\;\;\;\;\;\;\;\;\;\;\;\;\;\;\;\;\;\;\;\; \;\;\;\;\;\;\;\;\;\;\;\;\; & \\
A^{P}_{\rho^{+}}(a_{3},\cdots, a_{10})=(a_{4}+a_{10})f_{\rho}F_{1}(m_{\rho}^{2}). \;\;\;\;\;\;\;\;\;\;\;\;\;\;\;\;\;\;\;\;\;\;\;\;\;\;\;\;\;\;\;\;\;\;\;\;\;\;\;\;\;\;\;\;\;\;\;\;\;\;\;\;\;\;\;\;\;\;\; \;\;\;\;\;\;\;\;\;\;\;\;\;\;\;\;\;\;\;\;\; &
\end{eqnarray*}
Moreover, we can calculate the ratio between these two branching ratios, in which the uncertainty caused by many systematic errors is removed.  We define the ratio $R$  as: 
\begin{eqnarray}
R= \frac{BR(B^{0} \rightarrow \rho^{+} \pi^{-})}{BR(B^{+} \rightarrow \rho^{0} \pi^{+})},
\end{eqnarray}
and, without taking  into account the penguin contribution, one has,
\begin{eqnarray}
R=\frac{2 \Gamma_{B^{+}}}{ \Gamma_{B^{0}}} \bigg| \bigg( \frac{a_{1}}{a_{2}}+ \frac{f_{\pi}A_{0}(m_{\pi}^{2})}{f_{\rho}F_{1}(m_{\rho}^{2})} \bigg) \bigg(1+\frac{\tilde{\Pi}_{\rho \omega}}{(s_{\rho}-m_{\omega}^{2})+im_{\omega}\Gamma_{\omega}}\bigg) \bigg|^{-2}
\end{eqnarray}
\subsection{Numerical Results}
The latest experimental data from the CLEO collaboration~\cite{ref6} are:
\begin{eqnarray*}
BR(B^{+} \rightarrow \rho^{0}\pi^{+})=(10.4_{-3.4}^{+3.3} \pm 2.1) \times 10^{-6}, \;\;\;\; & & \\
BR(B^{0} \rightarrow \rho^{+}\pi^{-})=(27.6_{-7.4}^{+8.4} \pm 4.2) \times 10^{-6}, \;\;\;\;\;  & &\\
R=2.65 \pm 1.9.  \;\;\;\;\;\;\;\;\;\;\;\;\;\;\;\;\;\;\;\;\;\;\;\;\;\;\;\;\;\;\;\;\;\;\;\;\;\;\;\;\;\;\;\;\;\;\;\; & & 
\end{eqnarray*}
We have calculated the branching ratios  for $B^{0} \rightarrow \rho^{+}\pi^{-}$ and for $B^{+} \rightarrow \rho^{0}\pi^{+}$  for all models as a function of $N_{c}$. In  Figs.5 and 6,  we show the results for  models 1 and 2  in order to make  the dependence on form factors explicit.

The numerical results are very sensitive to uncertainties coming from  the experimental data. For the branching ratio   $B^{0} \rightarrow \rho^{+}\pi^{-}$  (Fig.5), we have a large range of values of $N_{c}$ and the  CKM matrix elements over  which  the  theoretical results are consistent with the experimental data from CLEO. However,  all models do not  give the same result: models 2 and 4  are very close to the experimental data for a large range of $N_{c}$, whereas models 1,3  and 5 are not. The reason is still the magnitude  of the form factors. As a result,  we have to exclude models 1,3  and 5 because  their form factors are too small. 

If we consider numerical results for branching ratio  $B^{+} \rightarrow \rho^{0} \pi^{+}$  (Fig.6), it appears that all models are consistent with the experimental data for a large range of $N_{c}$. The effect of $\rho-\omega$ mixing (included in our  calculations) on the branching ratio $B^{+} \rightarrow \rho^{0} \pi^{+}$ is around $30\%$. Numerical results for models 1, 3, 5 and models  2, 4 are very close to each other. The difference between the two branching ratios   can be explained   by the fact that for the $B^{0} \rightarrow \rho^{+}\pi^{-}$  decay, the tree and penguin contributions are both proportional to only one form factor, $F_{1}(k^{2})$. Thus, this branching ratio is very  sensitive to the magnitude  of this form factor ($F_{1}(k^{2})$ is related  to $h_{1}=0.330$ or $0.625$ in models  (1,3)  and (2,4)  respectively). On the other hand, for the  decay $B^{+} \rightarrow \rho^{0} \pi^{+}$, both  $F_{1}(k^{2})$ and $A_{0}(k^{2})$ are included in the  tree and penguin amplitudes, and    this branching ratio is less sensitive to the  magnitude of  the form factors. 

If we look at the ratio $R$ between these  two branching ratios,  $BR(B^{+}\rightarrow \rho^{0} \pi^{+})$ and $BR(B^{0}\rightarrow \rho^{+} \pi^{-})$  -- shown in Fig.7 -- the results indicate   that $R$ is very sensitive to the magnitude  of the  form factors,  and that there is  a large difference between models 1, 3, 5 and models 2 and  4. We investigated the ratio $R$ for the  limiting  CKM matrix elements as a function of $N_{c}$, finding   that $R$ is consistent with the experimental data  over  the range   $N_{c}$: $0.98(0.94)< N_{c}< 2.01(1.95)$,  (The values outside(inside) brackets correspond to the choice $q^{2}/m_{b}^{2}=0.3(0.5)$). It should be noted that $R$, in particular,  is not very  sensitive  to the   CKM matrix elements. The small difference which does appear, comes from the penguin contributions (which may be neglected). If we just take into account the tree contributions in our calculations, $R$ is clearly  independent of the  CKM matrix elements (Eq.42).

From  a comparison of the  numerical results and the experimental data, we can extract a range  of  $N_{c}$, within   which  all  results are consistent. In Table 3, we have summarized the allowed  range of  $N_{c}$ for  $B^{+} \rightarrow \rho^{0}\pi^{+}$, $B^{0} \rightarrow \rho^{+}\pi^{-}$  and $R$, for models 1, 2, 3, 4 and  5  according to  various choices of  the CKM matrix elements. To determine the  best range of $N_{c}$, we have to find some  intersection of the values of $N_{c}$ for each model and for each set of  CKM matrix elements, for which  the   theoretical and experimental results are consistent. This is possible and  the  results are shown in  Table 4. In our study, it seems  better to  use the range intersection   $\{N_{c}\}_{B^{+}} \cap \{N_{c}\}_{R}$ than   $\{N_{c}\}_{B^{0}} \cap \{N_{c}\}_{B^{+}}$, for fixing the final interval   $N_{c}$, since  the experimental  uncertainties  are smaller  in the former case, and since we are working to the first order of isospin violation ($\rho-\omega$ mixing). Finally, after excluding   models  1,3  and 5,  which are not consistent with all the  experimental data, we are  able to fix the upper and lower limit of the range of $N_{c}$, using the limiting values of the   CKM matrix elements (Table 5). We find that $N_{c}$ should be in the range $0.98(0.94)< N_{c}< 2.01(1.95)$ where $N_{cmin}$ and $N_{cmax}$ correspond to   $(\rho_{min},\eta_{min})$ and $(\rho_{max},\eta_{max})$  respectively.

\section{Summary and discussion}
The first aim of the present work was   to compare our theoretical results  with the  latest  experimental data from the CLEO collaboration for the branching ratios    $B^{+} \rightarrow \rho^{0}\pi^{+}$ and  $B^{0} \rightarrow \rho^{+}\pi^{-}$. Our next aim was to study  direct CP violation  for the decay $ B^{+} \rightarrow \rho^{0}(\omega) \pi^{+} \rightarrow \pi^{+} \pi^{-} \pi^{+}$,   with the inclusion  of $\rho-\omega$ mixing. The advantage of $\rho-\omega$ mixing is that the strong phase difference is large and rapidly varying   near  the $\omega$ resonance.  As a result the CP violating asymmetry, $a$,  has a maximum, $a_{max}$,  when the invariant mass of the $\pi^{+}\pi^{-}$ pair is in the vicinity of the $\omega$ resonance and $\sin\delta=+1$ at this point.

In the calculation of CP violating asymmetry parameters, we need the Wilson coefficients for the tree and penguin  operators at the scale $m_{b}$. We  worked with the renormalization scheme independent Wilson coefficients.  One of the major  uncertainties  is that  the hadronic matrix elements for both tree and penguin  operators  involve  nonperturbative QCD.  We have worked in the factorization approximation,  with $N_{c}$ treated as an effective parameter. Although one must have some doubts about factorization, it has been pointed out that  it may be quite reliable in energetic weak decays~\cite{ref22,ref23}.

 We have explicitly shown that the  CP violating asymmetry, $a$, is very  sensitive to the CKM matrix elements and the magnitude of the form factors, and we have determined a range  for the maximum asymmetry, $a_{max}$, as a function of the parameter $N_{c}$, the limits of CKM matrix elements and the choice of $k^{2}/m_{b}^{2}=0.3(0.5)$. From all the  models investigated, we found that CP violating asymmetry, $a_{max}$, varies from $-24\%(-19\%)$ to $-59\%(-48\%)$. We stressed that the ratio between the asymmetries  associated with the limiting values of CKM matrix elements would be mainly determined  by $\eta$. Moreover, we also stressed that without $\rho-\omega$ mixing, we cannot  have a  large CP violating asymmetry, a,  since   $a$ is proportional to both $\sin \delta$ and $r$.  Even though $\sin \delta$ is large around  $N_{c}=1$, $r$ is very small. As a result, we find a very small value for the CP violation in the decay  $B^{\pm} \rightarrow \rho^{0}\pi^{\pm}$  (of the order of a few percent)  without mixing. Once mixing is included, the sign of $\sin\delta$ is positive for  $N_{c}:0.98(0.94)< N_{c}< 2.01(1.95)$. Indeed, at the $\pi^{+}\pi^{-}$ invariant mass where the asymmetry, $a$,  is  maximum, $\sin\delta=+1$, independent of the parameters used. Thus, by measuring $a$, we can erase the phase uncertainty mod($\pi$) in the determination of the CKM angle $\alpha$ which arises from the conventional determination of $\sin2\alpha$.

The theoretical results for the  branching ratios for $B^{+} \rightarrow \rho^{0}\pi^{+}$ and  $B^{0} \rightarrow \rho^{+}\pi^{-}$,  were compared with  the experimental data from the CLEO collaboration~\cite{ref6}. These calculations show that it is  possible to have  theoretical results consistent with the experimental data without needing to invoke contributions from other  resonances~\cite{ref24,ref25}.  These data helped us to constrain   the magnitude of the various  form factors needed in the theoretical calculations  of B  decays\footnote{We note that  BABAR  reported preliminary branching ratios for this channel after this paper was prepared~\cite{ref26}. These results are consistent with the CLEO values.}.  We determined a range of value of  $N_{c}$,
 $0.98(0.94)< N_{c}< 2.01(1.95)$, inside of which  the  experimental data and  the theoretical calculations  are consistent for models 2  and 4.

 We will need more accurate data in the future to further decrease  the  uncertainties in the calculation. If we can use both the  CP violating  asymmetry  and the  branching ratios, with smaller uncertainties, we expect to be able to   determine the CKM matrix elements more precisely.  At the very least, it appears that one will be able to unambiguouly determine the sign of $\sin \alpha$ and hence, remove the well known discrete uncertainties in $\alpha$ associated with the fact that indirect CP violation determines only $\sin 2\alpha$. We expect that our predictions should provide useful guidance for future investigations and urge our experimental collegues to plan seriously to measure the rather dramatic direct CP violation predicted here.
\newline
\newline
%\vspace{5em}
{\bf Acknowledgments:}

This work was supported  in part by the Australian Research Council and the University of Adelaide.

\newpage

\newpage
%References

%
%
\newpage
{\bf \Large Figure Captions }
\newline
\newline
Fig.1 Asymmetry, $a$, for $k^{2}/m_{b}^{2}=0.3$, $ N_{c}=0.98(2.01)$  and limiting values of the CKM matrix elements for  model 1: solid line(dot line) for $N_{c}=0.98$ and max(min) CKM matrix elements.  Dashed line(dot dashed line) for $N_{c}=2.01$ and max(min) CKM matrix elements. 
\newline
\newline
Fig.2 Asymmetry, $a$, for $k^{2}/m_{b}^{2}=0.5$, $ N_{c}=0.94(1.95)$  and limiting values of the CKM matrix elements for  model 1: solid line(dot line) for $N_{c}=0.94$ and max(min) CKM matrix elements. Dashed line(dot dashed line) for $N_{c}=1.95$ and max(min) CKM matrix elements. 
\newline
\newline
Fig.3 Determination of the strong phase difference, $\sin\delta$, for $k^{2}/m_{B}^{2}=0.3(0.5)$ and for   model 1.   Solid  line(dot line) for  $\tilde{\Pi}_{\rho \omega}=(-3500;-300)$ (i.e. with $\rho - \omega$ mixing). Dot dashed line(dot dot dashed line) for  $\tilde{\Pi}_{\rho \omega}=(0;0)$, (i.e. with no $\rho - \omega$ mixing).  
\newline
\newline
Fig.4 Evolution of the ratio of penguin to tree amplitudes, $r$,   for $k^{2}/m_{B}^{2}=0.3(0.5)$, for limiting values of the CKM matrix elements $(\rho,\eta)$ max(min), for  $\tilde{\Pi}_{\rho \omega}=(-3500;-300)(0,0)$, (i.e. with(without) $\rho - \omega$  mixing) and for   model 1. Figure 4a (left): for $\tilde{\Pi}_{\rho \omega}=(0;0)$,  solid line(dot line) for $k^{2}/m_{B}^{2}=0.3$ and $(\rho,\eta)$ max(min). Dot dashed line(dot dot dashed line) for  $k^{2}/m_{B}^{2}=0.5$ and $(\rho,\eta)$ max(min). Figure 4b (right): same caption but for  $\tilde{\Pi}_{\rho \omega}=(-3500;-300)$. 
\newline
\newline
Fig.5 Branching ratio for $B^{0} \rightarrow \rho^{+}  \pi^{-}$ for models  $1(2)$, $k^{2}/m_{B}^{2}=0.3$ and limiting values of the  CKM matrix elements. Solid line(dot line) for  model 1 and max(min) CKM matrix elements. Dot dashed line(dot dot dashed line) for  model 2 and max(min) CKM matrix elements.  
\newline
\newline
Fig.6 Branching ratio for $B^{+} \rightarrow \rho^{0}\pi^{+}$ for models  $1(2)$, $k^{2}/m_{B}^{2}=0.3$ and limiting values of the  CKM matrix elements. Solid line(dot line) for  model 1  and max(min) CKM matrix elements. Dot dashed line(dot dot dashed line) for  model 2 and max(min) CKM matrix elements. 
\newline
\newline
Fig.7 Calculation  of the ratio of the two $\rho \pi$ branching ratios versus $N_{c}$  for models $1(2)$ and for limiting values of the CKM matrix elements: solid line(dot line) for   model 1 with max(min) CKM matrix elements.  Dot dashed line(dot dot dashed line) for   model 2 with max(min) CKM matrix elements. 
\newpage
{\bf \Large Tables }
\newline
\newline
Table 1 Maximum CP violating asymmetry, $a_{max}(\%)$,  for $B^{+} \rightarrow \pi^{+} \pi^{-} \pi^{+}$, for all models, limiting values of the  CKM matrix elements (upper and lower limit), and for $k^{2}/m_{b}^{2}=0.3(0.5)$. 
\newline
\newline
Table 2 Values of the CKM unitarity triangle for limiting values of the CKM matrix elements. 
\newline
\newline
Table 3 Summary of the range of values of $N_{c}$ which is determined from  the experimental data for various  models and input parameters. 
\newline
\newline
Table 4 Determination of the  intersection of the values of $N_{c}$ which are consistent with various subsets of the data   for all models and all sets of  CKM matrix elements. 
\newline
\newline
Table 5 Best range of  $N_{c}$ determined from  Table $4$ for $k^{2}/m_{b}^{2}=0.3(0.5)$.
\newline
\newline
\newpage
\begin{figure}[p]
\begin{center}
{\epsfsize=14.7in\epsfbox{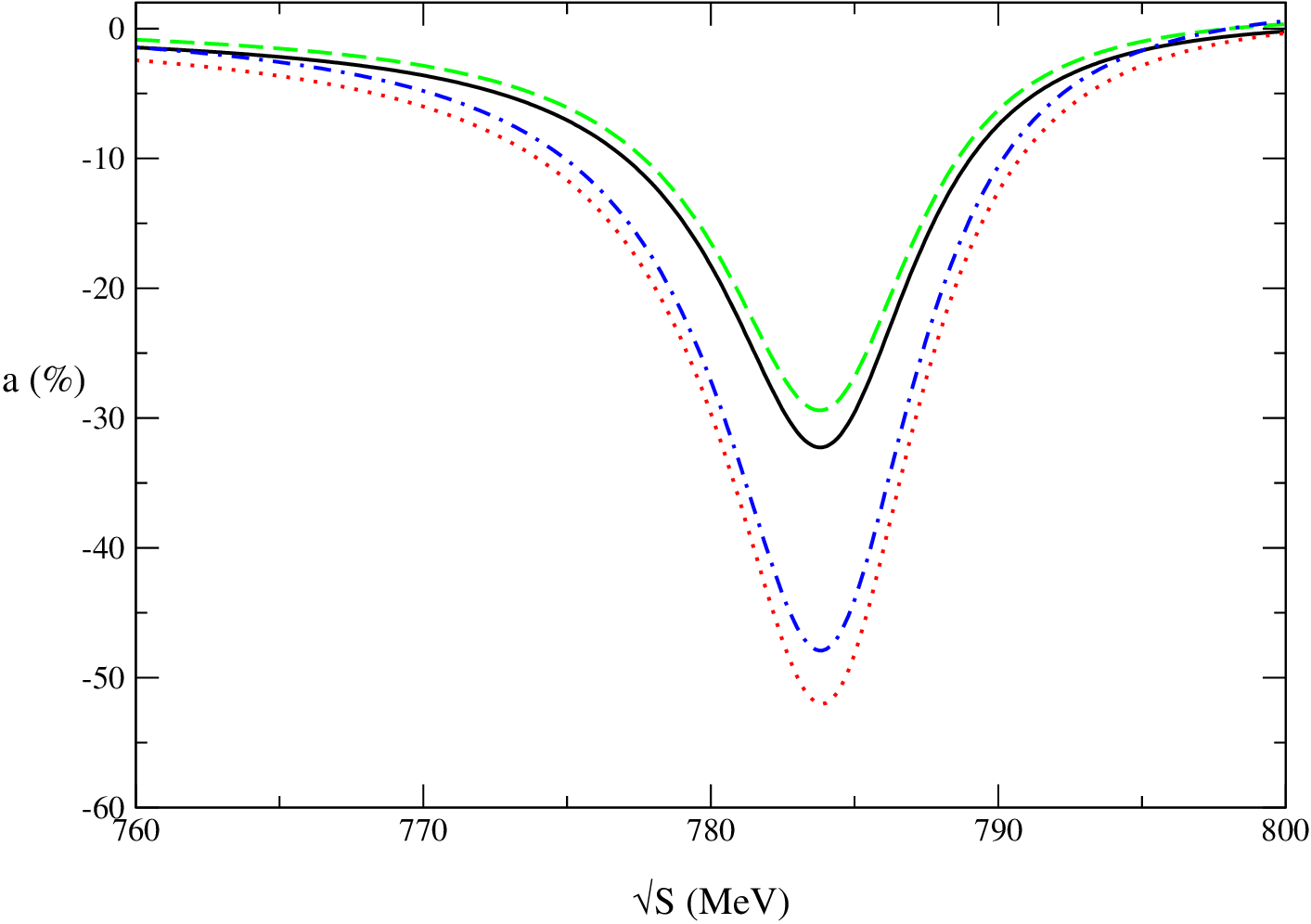}}
\caption{Asymmetry, $a$, for $k^{2}/m_{b}^{2}=0.3$, $ N_{c}=0.98(2.01)$  and limiting values of the CKM matrix elements for   model 1: solid line(dot line) for $N_{c}=0.98$ and max(min) CKM matrix elements.  Dashed line(dot dashed line) for $N_{c}=2.01$ and max(min) CKM matrix elements.}
\vspace{4em}
{\epsfsize=14.7in\epsfbox{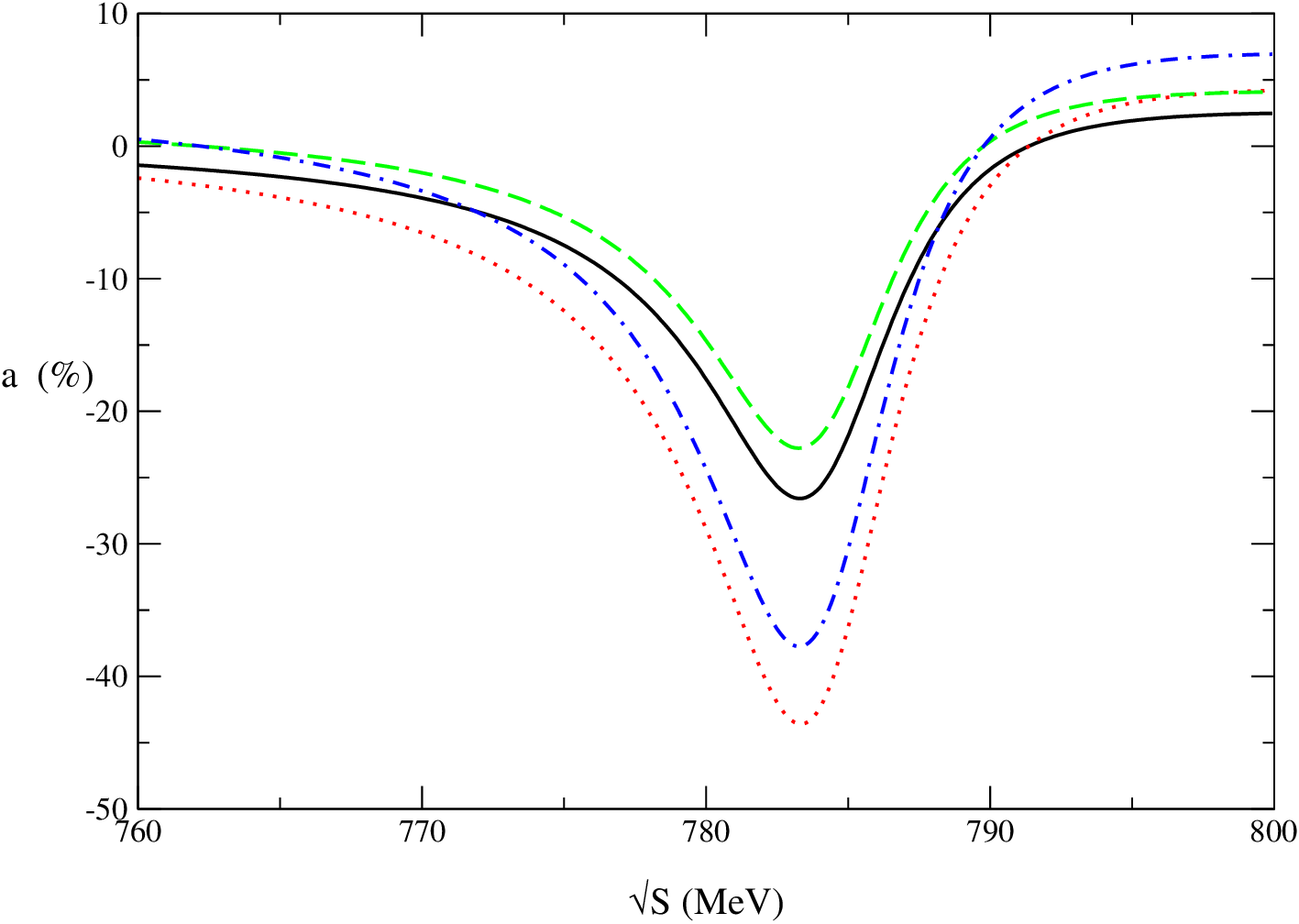}}
\caption{Asymmetry, $a$, for $k^{2}/m_{b}^{2}=0.5$, $ N_{c}=0.94(1.95)$  and limiting values of the  CKM matrix elements for  model 1: solid line(dot line) for $N_{c}=0.94$ and max(min) CKM matrix elements. Dashed line(dot dashed line) for $N_{c}=1.95$ and max(min) CKM matrix elements.}
\end{center}
\end{figure}
\newpage
\begin{figure}[p]
\begin{center}
{\epsfsize=14.7in\epsfbox{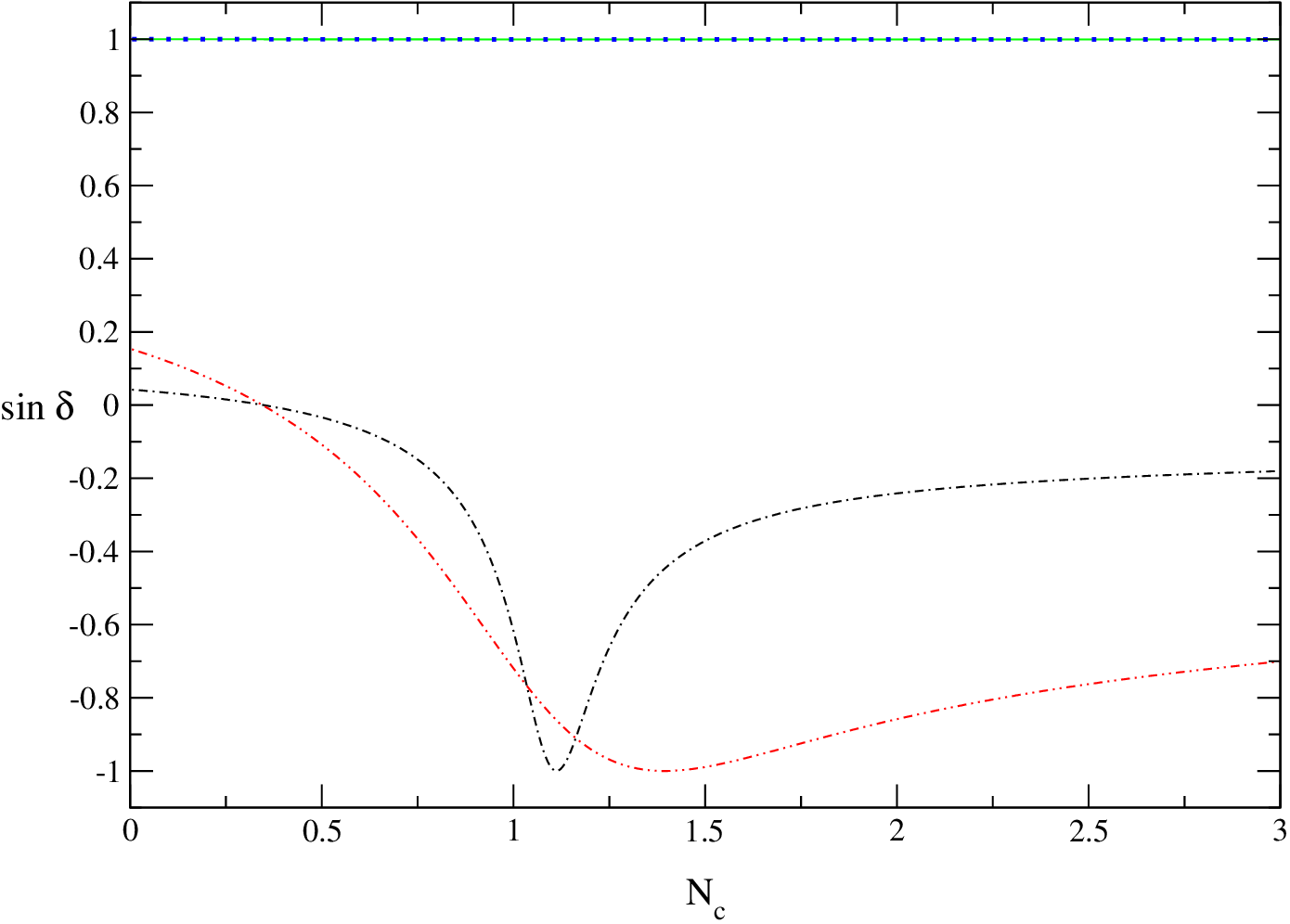}}
\caption{Determination of the strong phase difference, $\sin\delta$, for $k^{2}/m_{B}^{2}=0.3(0.5)$ and for  model 1.   The solid(dotted) line at $\sin \delta=+1$ corresponds the case    $\tilde{\Pi}_{\rho \omega}=(-3500;-300)$, where      $\rho - \omega$ mixing is included. The  dot dashed(dot dot dashed) line corresponds to  $\tilde{\Pi}_{\rho \omega}=(0;0)$, where   $\rho - \omega$ mixing is not included. }
\vspace{3.em}
{\epsfsize=14.6in\epsfbox{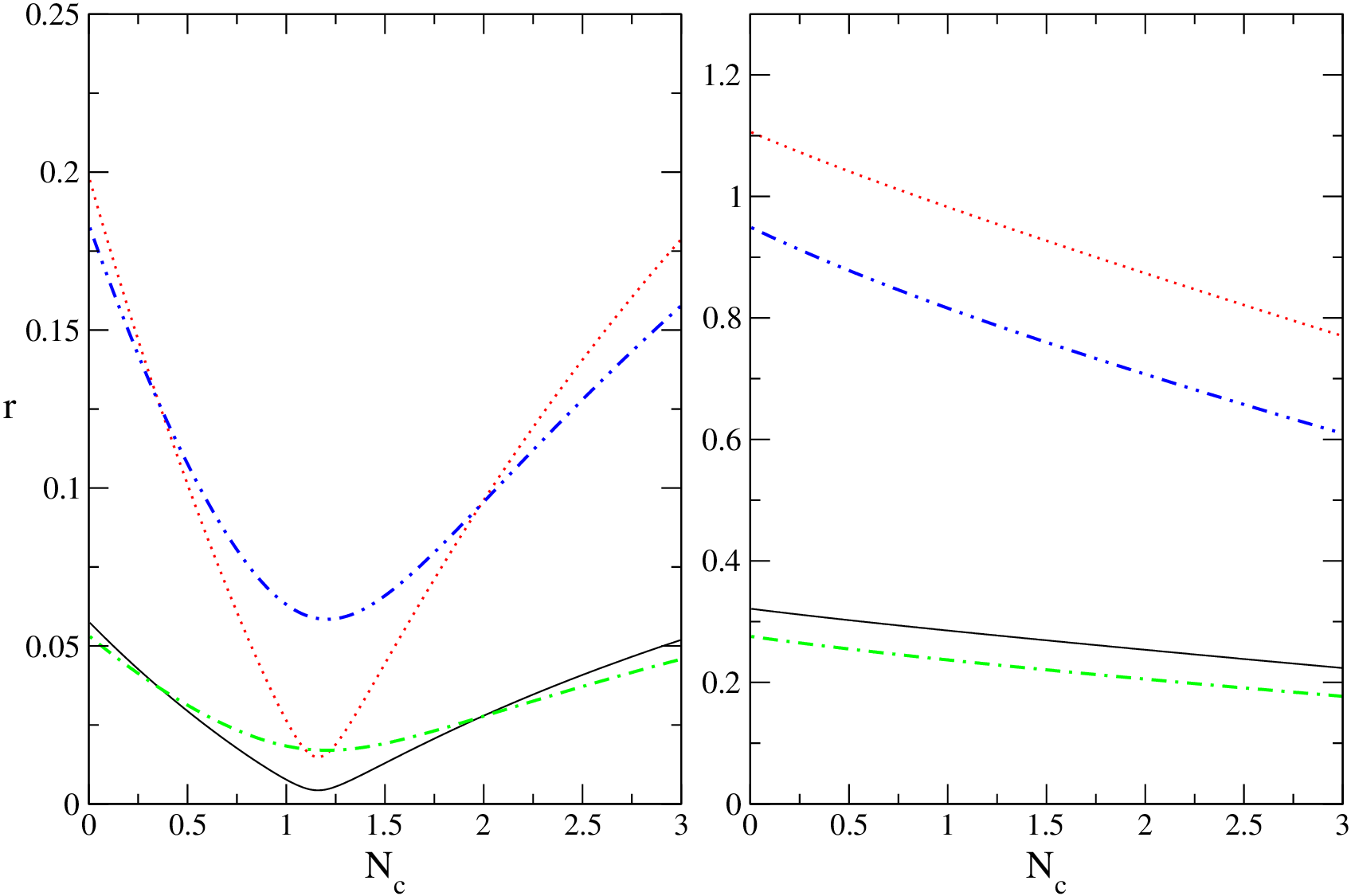}}
\caption{Evolution of the ratio of penguin to tree amplitudes, $r$, for $k^{2}/m_{B}^{2}=0.3(0.5)$, for limiting values of the CKM matrix elements $(\rho,\eta)$ max(min),  for  $\tilde{\Pi}_{\rho \omega}=(-3500;-300)(0,0)$, (i.e. with(without) $\rho - \omega$  mixing) and for   model 1. Figure 4a (left): for $\tilde{\Pi}_{\rho \omega}=(0;0)$,  solid line(dot line) for $k^{2}/m_{B}^{2}=0.3$ and $(\rho,\eta)$ max(min). Dot dashed line(dot dot dashed line) for  $k^{2}/m_{B}^{2}=0.5$ and $(\rho,\eta)$ max(min). Figure 4b (right): same caption but for  $\tilde{\Pi}_{\rho \omega}=(-3500;-300)$. }
\end{center}
\end{figure}
\newpage
\begin{figure}[p]
\begin{center}
{\epsfsize=14.7in\epsfbox{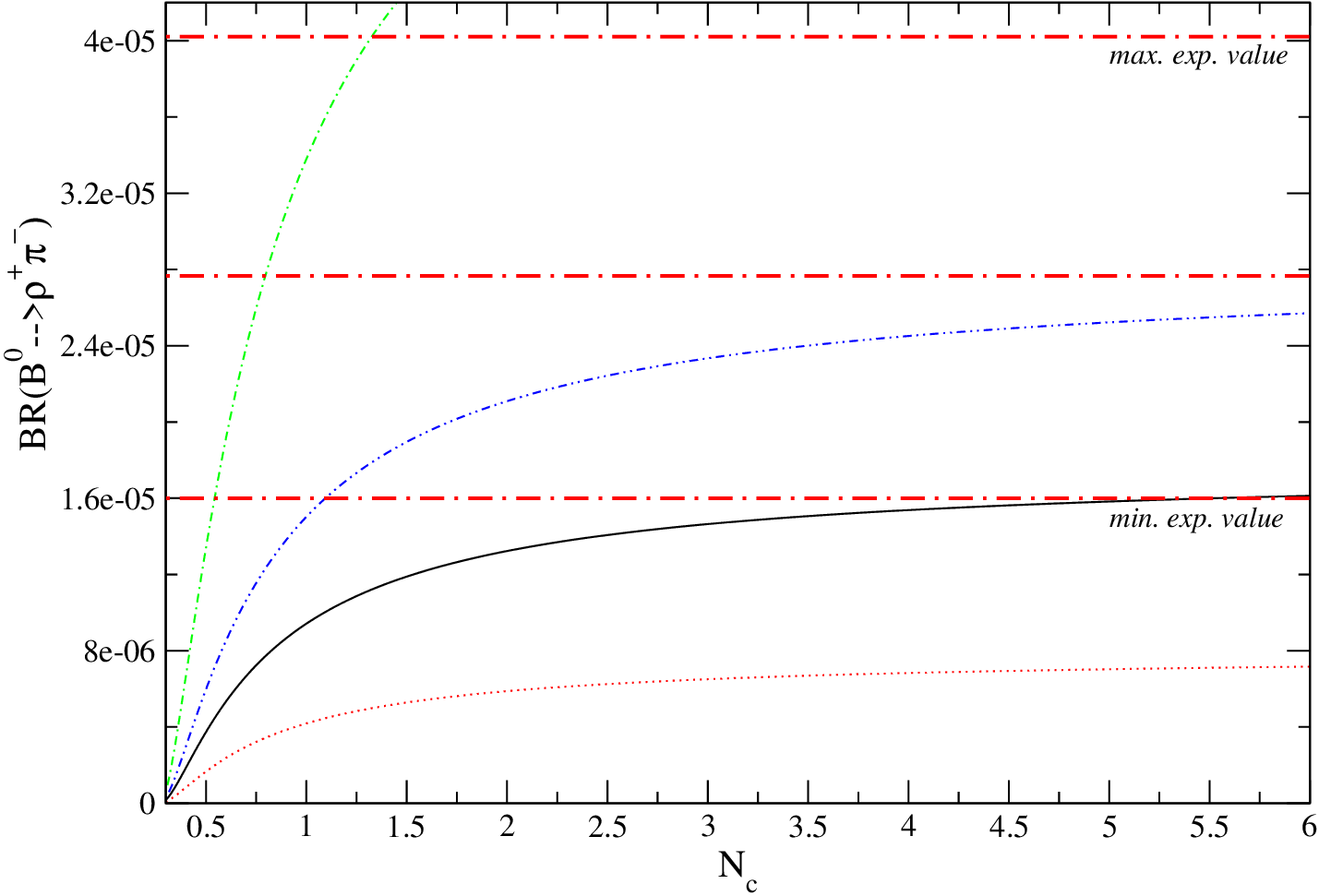}}
\caption{Branching ratio for $B^{0} \rightarrow \rho^{+}  \pi^{-}$ for models  $1(2)$, $k^{2}/m_{B}^{2}=0.3$ and limiting values of the  CKM matrix elements. Solid line(dot line) for  model 1 and max(min) CKM matrix elements. Dot dashed line(dot dot dashed line) for   model 2 and max(min) CKM matrix elements. }
\vspace{4em}
{\epsfsize=14.7in\epsfbox{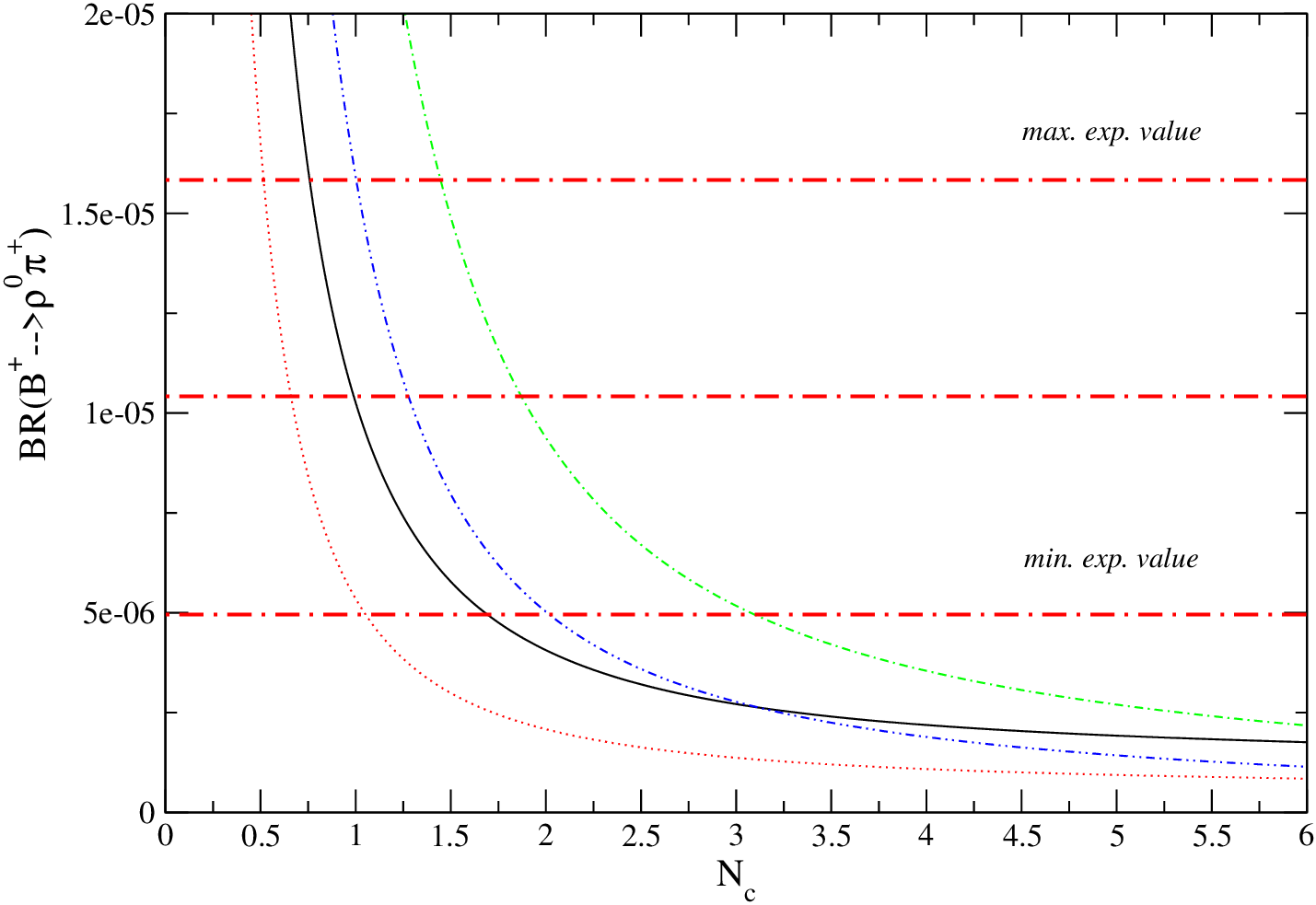}}
\caption{Branching ratio for $B^{+} \rightarrow \rho^{0}\pi^{+}$ for models  $1(2)$, $k^{2}/m_{B}^{2}=0.3$ and limiting values of the  CKM matrix elements. Solid line(dot line) for  model 1 and max(min) CKM matrix elements. Dot dashed line(dot dot dashed line) for  model 2 and max(min) CKM matrix elements. } 
\end{center}
\end{figure}
\newpage
\begin{figure}[p]
\begin{center}
{\epsfsize=15in\epsfbox{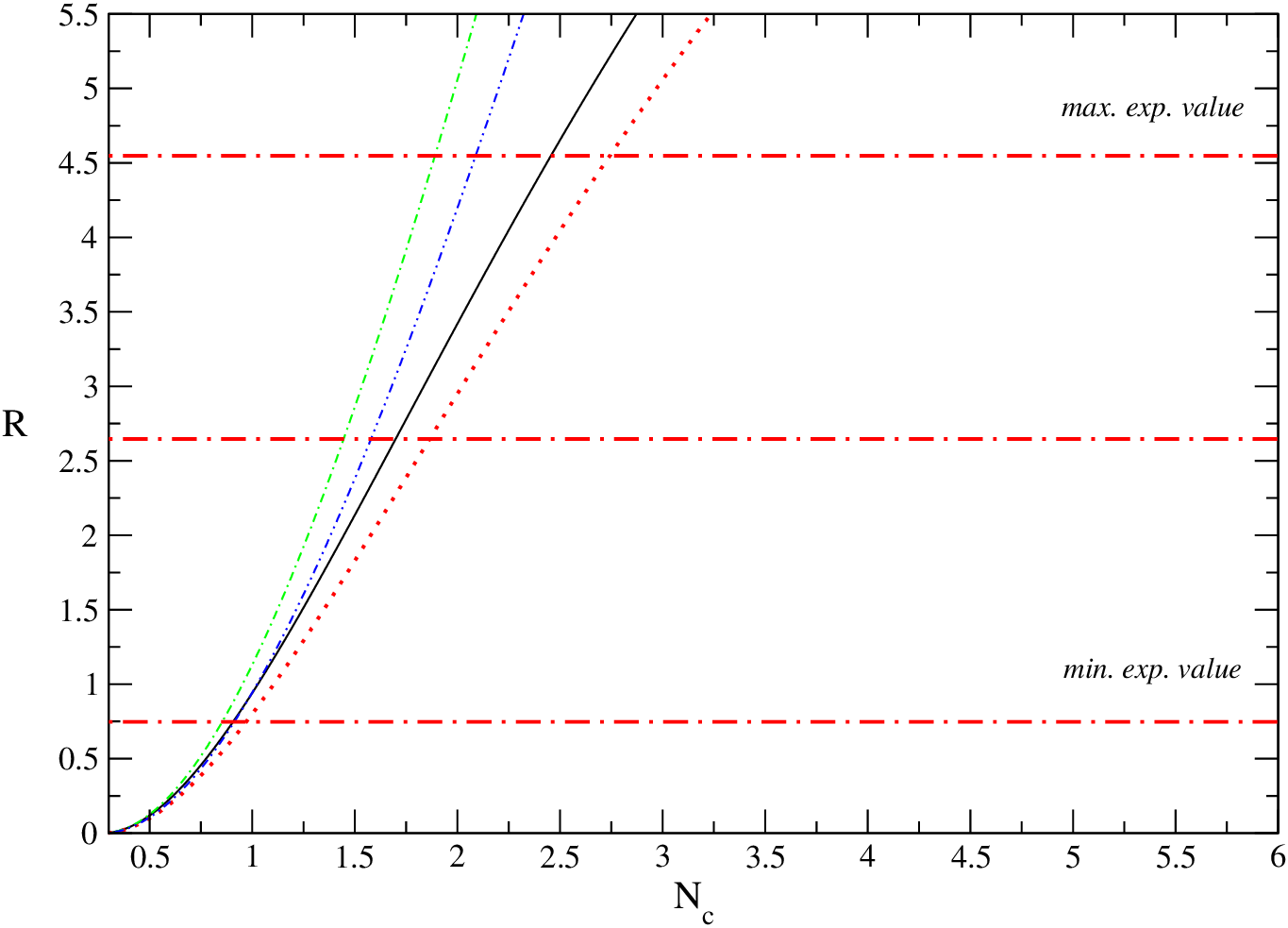}}
\caption{Calculation  of the ratio of two $\rho \pi$ branching ratios versus $N_{c}$  for models $1(2)$ and for limiting values of the CKM matrix elements: solid  line(dot line) for  model 1 with max(min) CKM matrix elements.  Dot  dashed line(dot dot  dashed line) for   model 2 with max(min) CKM matrix elements. }
\end{center}
\end{figure}
\newpage
\begin{table}[p]
\begin{center}
\begin{tabular}{ccc} \hline \hline    
                      & ${N_{c}}_{min}=0.98(0.94)$   & ${N_{c}}_{max}=2.01(1.95)$ \\
\hline
\hline
model $1$                                                                         \\
\hline
\hline
$ \rho_{max},\eta_{max}$   &        -33(-27)    &      -29(-23)                   \\
$ \rho_{min},\eta_{min}$   &        -52(-43)    &      -47(-37)                   \\
\hline                                  
\hline
model $2$                                                                        \\               
\hline
\hline
$ \rho_{max},\eta_{max}$   &        -36(-29)    &      -37(-28)                   \\
$ \rho_{min},\eta_{min}$   &        -57(-48)    &      -59(-46)                   \\
\hline  
\hline  
model $3$                                                                         \\
\hline
\hline
$ \rho_{max},\eta_{max}$   &        -32(-26)    &      -29(-23)                   \\ 
$ \rho_{min},\eta_{min}$   &        -51(-43)    &      -47(-37)                   \\
\hline  
\hline 
model $4$                                                                         \\
\hline
\hline
$ \rho_{max},\eta_{max}$   &        -36(-29)    &      -37(-28)                   \\ 
$ \rho_{min},\eta_{min}$   &        -57(-48)    &      -59(-46)                   \\ 
\hline  
\hline
model $5$                                                                         \\
\hline
\hline
$ \rho_{max},\eta_{max}$   &        -29(-24)    &      -24(-19)                   \\
$ \rho_{min},\eta_{min}$   &        -48(-40)    &      -39(-31)                   \\
\hline
\hline
\end{tabular}
\end{center}
\caption{Maximum CP violating asymmetry $a_{max}(\%)$ for $B^{+} \rightarrow \pi^{+} \pi^{-} \pi^{+}$, for all models, limiting values of the  CKM matrix elements (upper and lower limit), and for $k^{2}/m_{b}^{2}=0.3(0.5)$. }
\label{tab:fanfanfan}
\end{table}
\begin{table}[p]
\begin{center}
\begin{tabular}{ccc} \hline \hline    
                      & $(\rho,\eta)_{min}$   & $(\rho,\eta)_{max}$ \\

\hline
\hline
$\alpha$    &       $ 86^{o}02$    &     $ 89^{o}23$              \\
\hline
\hline
$\beta$  &       $ 19^{o}50$    &     $ 30^{o}64$                   \\
\hline
\hline
$\gamma$  &      $  74^{o}43$    &     $ 60^{o}11$                  \\ 
\hline  
\hline 
\end{tabular}
\end{center}
\caption{Values of the CKM unitarity triangle for limiting values of the CKM matrix elements. }
\label{tab:fanfanfanfan}
\end{table}

\newpage
\begin{table}[p]
\begin{center}
\begin{tabular}{cccc} \hline \hline  \hline       
                         & $B^{+}$              & $B^{0}$              &  $R$                   \\
\hline \hline  \hline 
 model $1$               &                      &                      &                        \\
\hline \hline 
$ \rho_{max},\eta_{max}$ & 0.76;1.69(0.73;1.62) & 5.50; $\star \star$ ( -- ; -- ) & 0.92;2.57(0.90;2.52)                      \\
\hline
$ \rho_{min},\eta_{min}$ & 0.52;1.04(0.49;0.98) &  -- ; -- ( -- ; -- )           & 0.97;2.88(0.94;2.76)   \\
\hline
$ \rho_{max},\eta_{min}$ & 0.61;1.25(0.59;1.20) &  -- ; -- ( -- ; -- )           & 0.92;2.58(0.91;2.54)   \\
\hline
$ \rho_{min},\eta_{max}$ & 0.69;1.46(0.66;1.39) &  -- ; -- ( -- ; -- )           & 0.95;2.75(0.90;2.66)   \\
\hline
\hline 
 model $2$               &                      &                      &                        \\
\hline \hline 
$ \rho_{max},\eta_{max}$ & 1.44;3.06(1.40;2.95) & 0.54;1.33(0.54;1.38)                      & 0.86;1.89(0.84;1.86)   \\
\hline
$ \rho_{min},\eta_{min}$ & 1.00;2.01(0.96;1.90) & 1.10; $\star \star$ (1.15; $\star \star$ ) & 0.92;2.09(0.89;2.01)   \\
\hline
$ \rho_{max},\eta_{min}$ & 1.15;2.32(1.12;2.22) & 0.70; $\star \star$ (0.72; $\star \star$ ) & 0.87;1.89(0.85;1.86)   \\
\hline
$ \rho_{min},\eta_{max}$ & 1.32;2.78(1.25;2.60) & 0.63;2.77(0.62;3.12)                      & 0.90;2.00(0.84;1.94)   \\
\hline
\hline
 model $3$               &                      &                      &                        \\
\hline \hline 
$ \rho_{max},\eta_{max}$ & 0.74;1.65(0.72;1.60) &   -- ; -- ( -- ; -- ) & 0.92;2.65(0.92;2.60)   \\
\hline
$ \rho_{min},\eta_{min}$ & 0.51;1.02(0.49;0.98) &   -- ; -- ( -- ; -- ) & 0.97;2.95(0.94;2.85)   \\
\hline
$ \rho_{max},\eta_{min}$ & 0.60;1.22(0.57;1.19) &   -- ; -- ( -- ; -- ) & 0.93;2.66(0.92;2.61)   \\
\hline
$ \rho_{min},\eta_{max}$ & 0.67;1.43(0.65;1.37) &   -- ; -- ( -- ; -- ) & 0.92;2.79(0.92;2.71)   \\
\hline
\hline
 model $4$               &                      &                      &                        \\
\hline 
\hline 
$ \rho_{max},\eta_{max}$ & 1.41;3.04(1.36;2.92) & 0.56;1.44(0.57;1.52)                     & 0.86;1.91(0.85;1.87)   \\
\hline 
$ \rho_{min},\eta_{min}$ & 0.98;1.96(0.94;1.87) & 1.16; $\star \star$ (1.23; $\star \star$ ) & 0.90;2.10(0.89;2.03)   \\
\hline
$ \rho_{max},\eta_{min}$ & 1.14;2.29(1.10;2.21) & 0.72; $\star \star$ (0.74; $\star \star$ ) & 0.86;1.92(0.85;1.88)   \\
\hline
$ \rho_{min},\eta_{max}$ & 1.30;2.74(1.24;2.59) & 0.64;3.49(0.66;4.03)                     & 0.89;2.01(0.86;1.95)   \\
\hline
\hline
 model $5$               &                      &                      &                         \\
\hline 
\hline 
$ \rho_{max},\eta_{max}$ & 0.75;2.18(0.73;2.10) &  -- ; -- ( -- ; -- ) & 1.03; $\star \star$ (1.02; $\star \star$ )    \\
\hline
$ \rho_{min},\eta_{min}$ & 0.50;1.08(0.47;1.03) &  -- ; -- ( -- ; -- ) & 1.09; $\star \star$ (1.06; $\star \star$ )    \\
\hline
$ \rho_{max},\eta_{min}$ & 0.58;1.38(0.55;1.34) &  -- ; -- ( -- ; -- ) & 1.03; $\star \star$ (1.02; $\star \star$ )    \\
\hline
$ \rho_{min},\eta_{max}$ & 0.66;1.71(0.64;1.62) &  -- ; -- ( -- ; -- ) & 1.04; $\star \star$ (1.04; $\star \star$ )    \\
\hline
\hline
\end{tabular}
\end{center}
\caption{Summary of the range of values of $N_{c}$ which is determined  from  the experimental data for various  models and input parameters (numbers outside(inside) brackets are for $k^{2}/m_{b}^{2}=0.3(0.5)$). The notation:
\newline
({\rm number}  ; {\rm number}) means that there is a upper and lower limit for $N_{c}$. ({\rm number}  ; $\star \star$ ) means that there is no upper limit for  $N_{c}$ in the range $N_{c}$ [0;10]. ( -- ; -- ) means that there is no range of $N_{c}$ which is consistent with experimental data. }
\label{tab:ecu}
\end{table}
\newpage
\begin{table}[p]
\begin{center}
\begin{tabular}{cccc} \hline \hline  \hline    
& $\{N_{c}\}_{B^{+}} \cap \{N_{c}\}_{B^{0}}$    &  $\{N_{c}\}_{B^{+}} \cap \{N_{c}\}_{R}$ & $\{N_{c}\}_{B^{0}} \cap \{N_{c}\}_{R}$    \\
\hline \hline  \hline  
 model $1$               &                      &                      &                         \\
\hline 
\hline 
$ \rho_{max},\eta_{max}$ &   -- ( -- )          & 0.92;1.69(0.90;1.62) &  -- ( -- )             \\
\hline
$ \rho_{min},\eta_{min}$ &   -- ( -- )          & 0.97;1.04(0.94;0.98) &  -- ( -- )              \\
\hline
$ \rho_{max},\eta_{min}$ &   -- ( -- )          & 0.92;1.25(0.91;1.20) &   -- ( -- )             \\
\hline
$ \rho_{min},\eta_{max}$ &   -- ( -- )          & 0.95;1.46(0.90;1.39) &   -- ( -- )             \\
\hline
\hline
 model $2$               &                      &                      &                         \\
\hline 
\hline 
$ \rho_{max},\eta_{max}$ &   -- ( -- )          & 1.44;1.89(1.40;1.86) & 0.86;1.33(0.84;1.38)    \\
\hline
$ \rho_{min},\eta_{min}$ & 1.10;2.01(1.15;1.90) & 1.00;2.01(0.96;1.90) & 1.10;2.09(1.15;2.01)    \\
\hline
$ \rho_{max},\eta_{min}$ & 1.15;2.32(1.12;2.22) & 1.15;1.89(1.12;1.86) & 0.87;1.89(0.85;1.86)  \\
\hline
$ \rho_{min},\eta_{max}$ & 1.32;2.78(1.25;2.60) & 1.32;2.00(1.25;1.94) & 0.90;2.00(0.84;1.94)   \\
\hline 
\hline
 model $3$               &                      &                      &                         \\
\hline 
\hline 
$ \rho_{max},\eta_{max}$ &     -- ( -- )        & 0.92;1.65(0.92;1.60) &      -- ( -- )             \\
\hline
$ \rho_{min},\eta_{min}$ &     -- ( -- )        & 0.97;1.02(0.94;0.98) &      -- ( -- )              \\
\hline
$ \rho_{max},\eta_{min}$ &     -- ( -- )        & 0.93;1.22(0.92;1.19) &      -- ( -- )             \\
\hline
$ \rho_{min},\eta_{max}$ &     -- ( -- )        & 0.92;1.43(0.92;1.37) &      -- ( -- )              \\
\hline
\hline
 model $4$               &                      &                      &                         \\
\hline 
\hline 
$ \rho_{max},\eta_{max}$ & 1.41;1.44(1.36;1.52) & 1.41;1.91(1.36;1.87) & 0.86;1.44(0.85;1.52)   \\
\hline 
$ \rho_{min},\eta_{min}$ & 1.16;1.96(1.23;1.87) & 0.98;1.96(0.94;1.87) & 1.16;2.10(1.23;2.03)    \\
\hline
$ \rho_{max},\eta_{min}$ & 1.14;2.29(1.10;2.21) & 1.14;1.92(1.10;1.88) & 0.86;1.92(0.85;1.88)    \\
\hline
$ \rho_{min},\eta_{max}$ & 1.30;2.74(1.24;2.59) & 1.30;2.01(1.24;1.95) & 0.89;2.01(0.86;1.95)     \\
\hline
\hline
 model $5$               &                      &                      &                         \\
\hline 
\hline 
$ \rho_{max},\eta_{max}$ &     -- ( -- )        & 1.03;2.18(1.02;2.10) &    -- ( -- )            \\
\hline
$ \rho_{min},\eta_{min}$ &     -- ( -- )        &       -- ( -- )      &    -- ( -- )            \\
\hline
$ \rho_{max},\eta_{min}$ &     -- ( -- )        & 1.03;1.38(1.02;1.34) &    -- ( -- )            \\
\hline
$ \rho_{min},\eta_{max}$ &     -- ( -- )        & 1.04;1.71(1.04;1.62) &    -- ( -- )            \\
\hline
\hline
\end{tabular}
\end{center}
\caption{Determination of the  intersection of the values of $N_{c}$ which are consistent with various subsets of the data   for all models and all sets of  CKM matrix elements (numbers outside(inside) brackets are for $k^{2}/m_{b}^{2}=0.3(0.5)$). The notation:
\newline
 -- ( -- ) means that  no common range of $N_{c}$ can be extracted from the  data.}
\label{tab:fan}
\end{table}
\begin{table}[p]
\begin{center}
\begin{tabular}{ccc} \hline \hline  \hline    
                        &   $\left\{N_{c}\right\}$ with mixing &  $\left\{N_{c}\right\}$ without  mixing    \\
\hline \hline  \hline  
model $2$               & 1.00;2.01(0.96;1.94)                  &     0.85;1.74(0.85;1.74)                                       \\
model $4$               & 0.98;2.01(0.94;1.95)                  &     0.84;1.76(0.84;1.75)
          \\   

\hline
\hline
\hline
maximum range           & 0.98;2.01(0.94;1.95)                  &     0.84;1.76(0.84;1.75)                                      \\
minimum range           & 1.00;2.01(0.96;1.94)                  &     0.85;1.74(0.85;1.74)                                     \\
\hline
\hline
\end{tabular}
\end{center}
\caption{Best range of  $N_{c}$ determined from  Table $4$ for $k^{2}/m_{b}^{2}=0.3(0.5)$. One takes the maximum interval of $N_{c}$, from Table 4, for each model (2,4). To determine the maximum(minimum) range, one consideres all models (2,4) and the largest(smallest) range of $N_{c}$.  In comparison, we show the range of $N_{c}$ determined without $\rho - \omega$ mixing. }
\label{tab:fanfan}
\end{table}
\end{document}